\documentstyle[12pt]{article}
\begin{document}

\tolerance=5000

\def\pp{{\, \mid \hskip -1.5mm =}}
\def\cL{{\cal L}}
\def\be{\begin{equation}}
\def\ee{\end{equation}}
\def\bea{\begin{eqnarray}}
\def\eea{\end{eqnarray}}
\def\tr{{\rm tr}\, }
\def\nn{\nonumber \\}
\def\gd{g^\dagger}
\def\e{{\rm e}}
\newcommand{\inv}[1]{\left[#1\right]_{\mbox{inv}}}
\newcommand\DS{D \hskip -3mm / \ }
\def\ds{\left( 1 + {M \over \lambda}
\e^{\lambda(\sigma^- - \sigma^+ )}
\right)}
\def\cR{{\cal R}}
\def\cF{{\cal F}}
\def\cE{{\cal E}^{-1}}
\def\e{{\rm e}}

\  \hfill 
\begin{minipage}{3.5cm}
OCHA-PP-114 \\
NDA-FP-45 \\
February 1998 \\
\end{minipage}

\vfill

\begin{center}
{\large\bf  Quantum cosmology in the models of 2d and 4d dilatonic 
 supergravity with WZ matter}

\vfill

{\sc S. James GATES, Jr.}\footnote{gates@umdhep.umd.edu},  
{\sc Tomoko KADOYOSHI}$^{\heartsuit}$\footnote{kado@fs.cc.ocha.ac.jp} \\
{\sc Shin'ichi NOJIRI$^{\clubsuit}$}\footnote{nojiri@cc.nda.ac.jp} and 
{\sc Sergei D. ODINTSOV$^{\spadesuit}$}\footnote{
 odintsov@galois.univalle.edu.co
}

\vfill

{\sl Department of Physics, University of Maryland at 
College Park \\
College Park, MD 20742-4111, USA}

\vfill

{\sl $\heartsuit$ Department of Physics, 
Ochanomizu University \\
Otsuka, Bunkyou-ku Tokyo 112, JAPAN}

\vfill

{\sl $\clubsuit$ 
Department of Mathematics and Physics \\
National Defence Academy, 
Hashirimizu Yokosuka 239, JAPAN}

\vfill

{\sl $\spadesuit$ 
Tomsk Pedagogical University, 634041 Tomsk, RUSSIA \\
and \\
Dep.de Fisica, Universidad del Valle, 
AA25360, Cali, COLOMBIA \\
}

\vfill

\ 

\newpage

{\bf ABSTRACT}

\end{center}

{ We consider $N=1$ two-dimensional (2d) dilatonic supergravity (SG), 
2d dilatonic SG 
obtained by dimensional reduction from $N=1$ four-dimensional (4d) SG, 
$N=2$ 2d dilatonic SG 
 and string-inspired 4d dilatonic SG. For all the theories, the corresponding 
action on a bosonic background is constructed and the interaction with $N$
(dilatonic) Wess-Zumino (WZ) multiplets is presented. Working in the 
large-$N$ approximation, 
it is enough to consider the trace anomaly induced effective action due to 
dilaton-coupled conformal matter as a quantum correction (for 2d models 
s-waves approximation is additionally used). The equations of motion 
for all such models with quantum corrections are written in a form convenient 
for numerical analysis. Their solutions are numerically investigated 
for 2d and 4d Friedmann-Robertson-Walker (FRW) or 4d Kantowski-Sacks 
Universes with a 
time-dependent dilaton 
via exponential dilaton coupling. The evolution of the corresponding 
quantum cosmological models is given for different choices of initial 
conditions and theory parameters. In most cases we find quantum singular 
Universes. Nevertheless, there are examples of Universe non-singular at 
early times. 
Hence, it looks unlikely that quantum matter back reaction on dilatonic
 background (at least in large $N$ 
approximation) may really help to solve the singularity problem. 
}

\ 


\newpage

\section{Introduction}

~~~It is widely known that models of two-dimensional (2d) dilaton gravity 
may  qualitatively describe the main properties of gravitational 
collapse and the evolution of the Universe taking into account 
quantum matter back reaction.  This is due to the fact that reduction 
of 4d gravitational theories with matter often leads to effective 
two-dimensional (dilatonic) gravity with dilaton coupled matter. In a 
recent work \cite{KNO}, we studied the quantum cosmology of 
four-dimensional (4d) reduced 
Einstein gravity with $N$ minimal scalars (working in large $N$ and 
$s$-waves approximations).  As a result of the reduction, one finds 2d 
dilatonic gravity with dilaton coupled scalars and quantum cosmology 
(of 4d Kantowski-Sacks form) where dilatonic coupling may be interpreted 
as a second scale factor of the singular, non-singular or big crunch type 
\cite{KNO}. 

It is quite interesting to generalize such a study for supergravities (SGs) 
which will be the purpose of this paper.  First, one can understand which 
different quantum SG may lead to which early Universe model in comparison
with purely bosonic versions. Second, some SGs represent low-energy 
effective actions of superstring theory and so such cosmologies (limiting 
to our approach) may possess some characteristic stringy cosmological 
features. Third, such study may help to select the ``best'' theories in 
the cosmological sense from among the existing variety of SGs. 

The present work is organized as follows. In the next section we review 
the construction of $N=1$ 2d dilatonic SG \cite{sdg} with $N$ scalar 
supermultiplets. Section three is devoted to $N=1$ 
Callan-Giddings-Harvey-Strominger (CGHS) SG with dilaton 
coupled scalar matter superfields. Working in the large $N$ approximation,
we take into account quantum matter effects and write the corresponding 
equations of motion (with non-zero matter fermions) on a bosonic background. 
The numerical study of such equations show 2d cosmologies where the Universe 
expands first and then shrinks, or where the Universe oscillates. In section 
four, we apply the same technique to study 2d dilatonic models with 
supermatter which is the reduced version from 4d $N=1$ SG with matter. 
Then two-dimensional cosmology with a dilaton may be interpreted as a 4d 
Kantowski-Sacks Universe. Numerical estimations show that it is singular 
4d Universe in the situations under discussion. In section five, 
we work with $N=2$ 2d dilatonic SG \cite{GATES} with dilaton-coupled 
matter. The numerical analysis of two-dimensional cosmologies with 
the back-reaction of quantum matter shows a qualitative difference 
from the $N=1$ SG case due to the presence of vector field in the bosonic 
background. Section six is devoted to string-inspired 4d dilatonic 
SG (its bosonic version) with $N$ dilaton coupled matter multiplets. Working 
in the large $N$ approximation and using the 4d anomaly induced action, we 
analytically and numerically investigate equations of motion for conformally 
flat 4d cosmologies. In the cases under consideration we have found  
singular or non-singular at early times or trivial cosmological solutions. 

\newpage

\section{General Action of $N=1$ Dilatonic Supergravity}

~~~The supersymmetric extension of the CGHS model \cite{12} was recently 
considered in \cite{NO}. This theory is a partial example of more general 
two-dimensional $N=1$ dilatonic supergravity \cite{sdg}.  The quantum 
corrections in the large $N$ approximation for such theories with matter have 
been discussed in \cite{sdg}. Using such quantum corrections one can also 
study Hawking radiation in 2d dilaton supergravity.  In this section, we 
briefly review the general action  using the component formulation of 
ref.\cite{HUY}.\footnote{The conventions and notations are given as follows
\begin{itemize}
\item signature
\[
\eta^{ab}=\delta^{ab}=\left(\begin{array}{cc}
1 & 0 \\
0 & 1 \end{array}\right)
\]
\item gamma matrices
\begin{eqnarray*}
\gamma^a\gamma^b&=&
\delta^{ab}+i\epsilon^{ab}\gamma_5 \ , \\
\sigma_{ab}&\equiv& {1 \over 4}[\gamma_a, \gamma_b]
={i \over 2}\epsilon_{ab}\gamma_5\ .
\end{eqnarray*}
\item charge conjugation matrix $C$
\begin{eqnarray*}
&&C\gamma_a C^{-1}=-\gamma_a^T \ , \nn
&& C=C^{-1}= -C^T \ , \\
&& \bar\psi=-\psi^T C\ .
\end{eqnarray*}
Here the index ${^T }$ means transverse.
\item Majorana spinor
\[
\psi=\psi^c\equiv C\bar\psi^T\ .
\]
\item Levi-Civita tensor
\begin{eqnarray*}
&&\epsilon^{12}=\epsilon_{12}=1\ ,
\ \ \epsilon^{ab}=-\epsilon^{ba}\ , 
\ \ \epsilon_{ab}=-\epsilon_{ba}\ , \\
&& \epsilon^{\mu\nu}=e\, e_a^\mu
e_b^\nu \epsilon^{ab} 
\ \ \epsilon_{\mu\nu}=e^{-1}\, e^a_\mu e^b_\nu 
\epsilon_{ab} \ .
\end{eqnarray*}
\end{itemize}}
In $N=1$ dilatonic supergravity  in this paper 
all the scalar fields are real and 
all the spinor fields are Majorana spinors.

The general $N=1$ action of 2d dilatonic supergravity is given in terms 
of general functions of the dilaton $C(\phi)$, $Z(\phi)$, $f(\phi)$ and 
$V(\phi)$ as follows
\bea
\label{lag}
&&\cL=-\inv{C(\Phi)\otimes W} \nn
&& +{1 \over 2}
\inv{\Phi\otimes\Phi\otimes T_P(Z(\Phi))} 
-\inv{Z(\Phi)\otimes\Phi \otimes
T_P(\Phi)} \nn
&& +\sum_{i=1}^N
\left\{{1 \over 2}
\inv{\Sigma_i\otimes\Sigma_i\otimes T_P(f(\Phi))}
-\inv{f(\Phi)\otimes \Sigma_i\otimes T_P(\Sigma_i)}
\right\} \nn
&& +\inv{V(\Phi)}\ .
\eea
Here $\Phi=(\phi,\chi,F)$ is a dilaton multiplet\footnote{The multiplet 
containing $C(\phi)$, for example, is given by $(C(\phi), C'(\phi)\chi, 
C'(\phi)F -{1 \over 2}C''(\phi)\bar\chi\chi)$. } and $\Sigma_i=(a_i,
\xi_i,G_i)$ is a matter multiplet.  These multiplets have the conformal 
weight $\lambda=0$.  $W$ is the curvature multiplet which is given by 
\be
\label{Wm}
W=\left(S,\eta,-S^2-{1 \over 2}R-{1 \over 2}
\bar\psi^\mu\gamma^\nu\psi_{\mu\nu}
+{1 \over 4}\bar\psi^\mu\psi_\mu\right)\ .
\ee
Here $R$ is the scalar curvature and 
\bea
\label{eta}
\eta&=&-{1 \over 2}S\gamma^\mu\psi_\mu 
+{i \over 2}e^{-1}\epsilon^{\mu\nu}\gamma_5
\psi_{\mu\nu} \ , \\
\label{psimn}
\psi_{\mu\nu}&=&D_\mu\psi_\nu - D_\nu\psi_\mu \ , \nn
D_\mu\psi_\nu&=&\left(\partial_\mu 
-{1 \over 2}\omega_\mu\gamma_5\right)\psi_\nu \ , \nn
\omega_\mu&=&-ie^{-1}e_{a\mu}\epsilon^{\lambda\nu}
\partial_\lambda e^a_\nu
-{1 \over 2}\bar\psi_\mu\gamma_5\gamma^\lambda
\psi_\lambda\ .
\eea
The curvature multiplet has the conformal weight $\lambda=1$. $T_P(Z)$ is 
called the kinetic multiplet for the multiplet $Z=(\varphi, \zeta, H)$ 
(in terms of superfield operators this corresponds to $(\nabla^2 
+ \lambda W) Z$) and when the multiplet $Z$ has the conformal weight 
$\lambda=0$, $T_P(Z)$ has the following form
\be
\label{kin}
T_P(Z)=(H, \DS\zeta, \Box\varphi)\ .
\ee
The kinetic multiplet 
$T_P(Z)$ has conformal weight $\lambda=1$.
The product of two multiplets 
$Z_k=(\varphi_k, \zeta_k, H_k)$ $(k=1,2)$ 
with the conformal weight $\lambda_k$ is defined by 
\be
\label{product}
Z_1\otimes Z_2 =(\varphi_1\varphi_2, 
\varphi_1\zeta_2 + \varphi_2\zeta_1, 
\varphi_1 H_2 + \varphi_2 H_1 - \bar\zeta_1\zeta_2)\ .
\ee
This is familiar in the language of superfields as the components 
of $Z_1 Z_2$. The invariant Lagrangian $\inv{Z}$ for the multiplet 
$Z$ is defined by
\be
\label{inv}
\inv{Z}=e\left[H+{1 \over 2}\bar\psi_\mu\gamma^\mu \zeta
+{1 \over 2}\varphi\bar\psi_\mu\sigma^{\mu\nu}\psi_\nu 
+S\varphi\right]\ ,
\ee
corresponding to the superaction ${\cal S} = \int d^2 \theta E^{-1}
Z$.

We consider only bosonic background fields below since this will be 
sufficient for the study of the cosmological problems under consideration.
On the bosonic background where the dilatino $\chi$ and the Rarita-Schwinger 
fields vanish, one can show that the gravity and dilaton part of the 
Lagrangian have the following forms:
\bea
\label{gdlag}
\inv{C(\Phi)\otimes W}  &\sim& 
e\left[-C(\phi)\left(S^2 +{1 \over 2}R\right)
-C'(\phi)FS\right] \ , \nn
\inv{\Phi\otimes\Phi\otimes T_P(Z(\Phi))} &\sim& 
e\left[\phi^2\tilde\Box(Z(\phi)) 
+ 2Z'(\phi)\phi F^2\right] \ , \nn
\inv{Z(\Phi)\otimes\Phi \otimes T_P(\Phi)} &\sim& 
e\left[Z(\phi)\phi\tilde\Box\phi 
+ Z'(\phi)\phi F^2 + Z(\phi)F^2 \right] \ , \nn
\inv{V(\Phi)} &\sim& e\left[ V'(\phi)F  + SV(\phi)\right]\ .
\eea
For matter part we obtain
\bea
\label{lag2}
&& \sum_{i=1}^N
\left\{{1 \over 2}
\inv{\Sigma_i\otimes\Sigma_i\otimes T_P(f(\Phi))}
-\inv{f(\Phi)\otimes \Sigma_i\otimes T_P(\Sigma_i)}\right\} \nn
&& \sim e f(\phi)\sum_{i=1}^N
(g^{\mu\nu}\partial_\mu a_i \partial_\nu a_i
+\bar\xi_i\gamma^\mu\partial_\mu\xi_i 
-f(\phi)G_i^2) \nn
&& + \ \mbox{total divergence terms}\ .
\eea
In eq.(\ref{gdlag}) and (\ref{lag2}), ``$\sim$'' means that we neglect 
the terms containing fermionic fields except the kinetic term of $\xi_i$.
Here we have used the fact that
\be
\label{maj}
\bar\xi_i \gamma_5 \xi=0
\ee
for the Majorana spinors $\xi_i$.
Using equations of motion with respect to the auxilliary fields $S$, 
$F$, $G_i$, on the bosonic background one can show that 
\bea
\label{bgaf}
S&=&{C'(\phi)V'(\phi) - 2V(\phi)Z(\phi) 
\over {C'}^2(\phi) + 4C(\phi) Z (\phi)} \ , \nn
F&=&{C'(\phi) V(\phi)+ 2 C(\phi)V'(\phi)
\over {C'}^2(\phi) + 4C(\phi) Z (\phi)} \ , \nn
G_i&=&0\ .
\eea
Especially for the supersymmetric extension \cite{NO} of 
the CGHS type model \cite{12} \be
\label{CGHS}
C(\phi)=2\e^{-2\phi}\ ,\ \ Z(\phi)=4\e^{-2\phi}\ ,
\ \ V(\phi)=4\e^{-2\phi}\ ,
\ee
we find
\be
\label{CGHSaf}
S=0\ ,\ \ F=-\lambda\ ,\ \ G_i=0\ .
\ee
We should note $\lambda^2>0$ in the supersymmetric 
theory or unitarity is broken. 

\section{$N=1$ Supersymmetric CGHS Model with Dilaton 
Coupled Matter}

~~~After integrating out the auxilliary fields as we have shown 
in the  previous section, the classical action of the CGHS type dilaton 
supergravity model with dilaton coupled supermatter in a purely bosonic 
background is given by
\bea
\label{bsSc}
S_c&=&
\int d^2x e \left[-{\e^{-2\phi} \over 2\pi}
\left\{R+4\partial_\mu\phi\partial^\mu\phi+ 4\lambda^2 \right.
\right. \nn
&& \left.\left. -{1 \over 2}\sum_{i=1}^N \left(
\partial_\mu a_i \partial^\mu a_i
+ \bar\xi_i \gamma^\mu D_\mu \xi_i\right)
\right\} \right]
\eea
As we are going to work in large $N$ approximation we may take 
into account only matter quantum effects. Then the bosonic part of  
the trace anomaly induced effective action \cite{trace,sdg} (for a pure 
dilaton coupled scalar, see also \cite{BH}) 
together with classical action has the following form\footnote{
The inverse of 
$\Box={1 \over \sqrt{-g}}\partial_\mu \sqrt{-g}g^{\mu\nu}\partial_\nu$ 
is defined by 
\[
\Box_x\left({1 \over \Box}(x,y)\right)=\sqrt{-g}\delta^4(x-y)\ .
\]
Here $x$ and $y$ are the cordinates of the space-time and the subscript $x$ 
of $\Box$ means the derivative with respect to $x$. In the conformal gauge
(\ref{cg}), 
the ambiguity of the boundary condition is absorbed into the function 
$t^\pm(x^\pm)$ which appears in (\ref{eqnpp}). }
\bea
\label{bsS2}
S&=&S_c+W \nn
&=& \int d^2x e \biggl[-{\e^{-2\phi} \over 2\pi}
\left\{R+4\partial_\mu\phi\partial^\mu\phi
-{1 \over 2}\sum_{i=1}^N \left(
\partial_\mu a_i \partial^\mu a_i 
+ \bar\xi_i \gamma^\mu D_\mu \xi_i\right) 
+ 4\lambda^2 \right\} \nn
&& -{1 \over 2\pi}\left\{ {N \over 32}R{1 \over \Box}R
- {N \over 4} \partial_\mu\phi\partial^\mu\phi 
{1 \over \Box} R
+ {N \over 4}\phi R \right\}\biggr]\ .
\eea
In the (super)conformal gauge,
\be
\label{cg}
g_{\pm\mp}=-{1 \over 2}\e^{2\rho}\ ,\ \ 
g_{\pm\pm}=0
\ee
the equations of motion given by the variations of 
$g_{\pm\pm}$, $g_{+-}$, $\phi$ and $a_i$ have the following form:
\bea
\label{eqnpp}
0&=&T_{\pm\pm} \nn
&=&\e^{-2\phi}\left(4\partial_\pm \rho
\partial_\pm\phi - 2 \left(\partial_\pm\phi\right)^2 
\right) + {1 \over 2}\e^{-2\phi}\sum_{i=1}^N 
(\partial_\pm a_i)^2 + T^f_{\pm\pm} \nn
&& +{N \over 8}\left( \partial_\pm^2 \rho 
- \partial_\pm\rho \partial_\pm\rho \right) \nn
&& +{N \over 2} \left\{
\left( \partial_\pm\phi \partial_\pm\phi \right)
\rho+{1 \over 2}{\partial_\pm \over \partial_\mp}
\left( \partial_\pm\phi \partial_\mp\phi \right)\right\} \nn
&& -{N \over 4}\left\{ -2 \partial_\pm \rho \partial_\pm \phi 
+\partial_\pm^2 \phi \right\} + t^\pm(x^\pm) \\
\label{req}
0&=&T_{\pm\mp} \nn
&=&\e^{-2\phi}\left(2\partial_+
\partial_- \phi -4 \partial_+\phi\partial_-\phi 
- \lambda^2 \e^{2\rho}\right) + T^f_{+-} \nn
&& -{N \over 8}\partial_+\partial_- \rho
-{N \over 4}\partial_+ \phi \partial_-  \phi 
+{N \over 4}\partial_+\partial_-\phi \\
\label{eqtp}
0&=& \e^{-2\phi}\biggl(-4\partial_+
\partial_- \phi +4 \partial_+\phi\partial_-\phi 
+2\partial_+ \partial_- \rho \nn
&& - {1 \over 2}\sum_{i=1}^N  \partial_+ a_i \partial_- a_i 
+ \lambda^2 \e^{2\rho}\biggr)- T^f_{+-} \nn
&& + \left\{
-{N \over 4}\partial_+(\rho \partial_- \phi)
-{N \over 4}\partial_-(\rho \partial_+ \phi)
-{N \over 4}\partial_+\partial_-\rho \right\} \\
\label{mat}
0&=& \partial_+(\e^{-2\phi}\partial_-\chi_i)
+\partial_-(\e^{-2\phi}\partial_+\chi_i)\ .
\eea
Here $t^\pm(x^\pm)$ is a function which is determined by 
the boundary condition and $T^f_{\mu\nu}$ is the energy-momentum 
tensor of the matter fermion defined by
\be
\label{Tf}
T^f_{\mu\nu}={1 \over 2}\e^{-2\phi}
\sum_{i=1}^N \left(\bar\xi_i\gamma_\mu\partial_\nu\xi_i
+ \bar\xi_i\gamma_\nu\partial_\mu\xi_i
- g_{\mu\nu}\bar\xi_i\gamma^\lambda\partial_\lambda
\xi_i\right)\ .
\ee
If we use the $\xi$-equation of motion, we find
\be
\label{Tfe}
T^f_{\pm\pm}=T^f_{\pm\pm}(x^\pm)\ ,\ \ T^f_{+-}=0\ .
\ee
Since we are presently considering the cosmological problem, we assume that 
all the fields depend only on time $t$ and replace $\partial_\pm\rightarrow 
{1 \over 2}\partial_t$.  Then Eq.(\ref{eqnpp}) tells $t^\pm$ and $T^f_{\pm
\pm}$ are constants: $t^\pm={N \over 4}t_0$ and $T^f_{\pm\pm}={N \over 4}T$.
Then Eqs.(\ref{eqnpp}), (\ref{req}), (\ref{eqtp}) 
and (\ref{mat})  can be rewritten as 
\bea
\label{eqnpp2}
0&=&\e^{-2\phi}\left(4\partial_t \rho
\partial_t \phi - 2 \left(\partial_t \phi\right)^2 
\right) + {1 \over 2} \e^{-2\phi}\sum_{i=1}^N 
(\partial_t  a_i)^2 + NT \nn
&& +{N \over 8}\left( \partial_t ^2 \rho 
- (\partial_t \rho)^2 \right) \nn
&& +{N \over 2} \left( \rho+{1 \over 2}\right)
\partial_t \phi \partial_t \phi \nn
&& -{N \over 4}\left\{ -2 \partial_t  \rho \partial_t  \phi 
+\partial_t ^2 \phi \right\} + N t_0 \\
\label{req2}
0&=&\e^{-2\phi}\left(2\partial_t^2
\phi -4 \partial_t\phi\partial_t\phi 
- 4\lambda^2 \e^{2\rho}\right) \nn
&& -{N \over 8}\partial_t^2 \rho
-{N \over 4}\partial_t  \phi \partial_t   \phi 
+{N \over 4}\partial_t^2 \phi \\
\label{eqtp2}
0&=& \e^{-2\phi}\left(-4\partial_t^2
\phi +4 (\partial_t \phi)^2 
+2\partial_t^2  \rho 
+ 4\lambda^2 \e^{2\rho} 
- {1 \over 2}\sum_{i=1}^N  (\partial_t  a_i)^2 \right)
\nn
&& + \left\{
-{N \over 2}\partial_t (\rho \partial_t  \phi)
-{N \over 4}\partial_t^2 \rho \right\} \\
\label{mat2}
0&=& \partial_t (\e^{-2\phi}\partial_t a_i)\ .
\eea
Eq.(\ref{mat2}) can be integrated to be $\partial_t a_i=A_i\e^{2\phi}$.
Here $A_i$ is a constant of the integration.
Note that $T$ can be absorbed into the redefinition of $t_0$
\be
\label{redeft0}
t_0 + T \rightarrow t_0\ .
\ee
In the following we use the redefinition of (\ref{redeft0}).

Solving Eqs.(\ref{eqnpp2}) and (\ref{req2}), we obtain
\bea
\label{ddp}
\partial_t^2 \phi &=& \left(3-{N \over 4}\rho\e^{2\phi}\right)
(\partial_t\phi)^2 
- \left(2 + {N \over 4}
\e^{2\phi}\right)\partial_t\rho\partial_t\phi 
+ {N \over 16} \e^{2\phi} (\partial_t\rho)^2 \nn 
&& - {N \over 4}\e^{4\phi}A^2 + 2\lambda^2\e^{2\rho}
- {N \over 2}\e^{2\phi}t_0
 \\
\label{ddr}
\partial_t^2 \rho &=& \left(1 + {N \over 8}\e^{2\phi}\right)
(\partial_t\rho)^2 
- {32 \over N}\e^{-2\phi} 
 \left(1 + {N \over 8}\e^{2\phi}\right)^2
\partial_t \rho \partial_t \phi \nn
&& + 4\left\{\left(1+{4 \over N}\e^{-2\phi} \right)
-\left(1 + {N \over 8}\e^{2\phi}\right)\rho\right\}
(\partial_t\phi)^2 \\
&& - 4\e^{2\phi}\left(1 + {N \over 8}\e^{2\phi}\right)A^2 
-8\left(1 + {N \over 8}\e^{2\phi}\right)t_0
+ 4\lambda^2 \e^{2\rho}\ .\nonumber
\eea
Here 
\be
\label{A2}
A^2 \equiv {1 \over N}\sum_{i=1}^N A_i^2\ .
\ee
Substituting (\ref{ddp}) and (\ref{ddr}) into (\ref{eqtp2}) we obtain
\bea
\label{t0}
t_0&=& -\left\{-16\e^{-2\phi} + 2N 
+ {N^2 \over 4}\e^{2\phi}\left(\rho + 1 \right)
\right\}^{-1} \nn
&& \times\Bigg[
\left\{{32 \over N}\e^{-4\phi} - (8\rho + 4)\e^{-2\phi}
- N\left(1+{\rho \over 2}\right)
+ {N^2 \over 8}\e^{2\phi}\left(\rho + 1 \right)\rho
\right\}(\partial_t\phi)^2 \nn
&& + \left\{ - {64 \over N}\e^{-4\phi} 
+N\left(\rho + {3 \over 2} \right)
+ {N^2 \over 8}\e^{2\phi}\left(\rho + 1\right)
\right\}\partial_t\rho \partial_t\phi \nn
&& + \left\{2\e^{-2\phi} - {N \over 4} 
- {N^2 \over 32}\e^{2\phi}\left(\rho + 1\right)
\right\}(\partial_t\rho)^2 \nn
&& + \left\{ -8 + {N \over 2}\e^{2\phi} + {N^2 \over 8}\e^{4\phi}
\left(\rho + 1 \right)\right\}A^2  \nn
&& +\{ 4\e^{-2\phi} - N(\rho +1)\}\lambda^2\e^{2\rho}\Bigg]
\eea
Eq. (\ref{t0}) determines $t_0$ from the initial conditions 
for $\phi$, $\partial_t\phi$, $\rho$ and $\partial_t\rho$. 
Note that attempt to study quantum cosmology in CGHS-like model 
(working in s-wave approximation and sigma-model description 
with ad hoc choice of dilatonic potential) has been done in ref.\cite{KB}. 
In our impression, above approach is better for treating the dynamical 
dilaton effects. Moreover, it can be applied to wider class of models.

Equations (\ref{ddp}), (\ref{ddr}) and (\ref{t0}) 
can be solved numerically for several parameters
with the initial condition $\phi=\rho={d\phi \over dt}
={d\rho \over dt}=0$ at $t=0$.
Typical graphs are given in Figs.1 -- 4.
We also calculated the two-dimensional scalar curvature $R$, 
which is given by
\be
\label{sR}
R=8\e^{-2\rho}\partial_+\partial_-\rho \ .
\ee
In Fig.1, both  the conformal factor $\rho$ and dilaton field $\phi$ 
increase monotonically with time and there is a curvature singularity 
in a finite conformal time.  We should note that there was not found 
a solution where $\rho$ monotonically increases with time in the 
bosonic model \cite{KNO}.  If $\rho$ diverges near the singularity, the 
singularity will appear in the infinite future of the cosmological time
defined by
\be
\label{ct}
d\hat t = \e^\rho dt\ .
\ee
Note $g_{tt}=1$ when we use the cosmological time $\hat t$.  It should 
be also noted that there will not  appear any singularities in the 
two-dimensional scalar curvature in the solutions of Figs.2, 3 and 4.
In Fig.2, $\phi$ increases monotonically in time but $\rho$ increases 
first and decreases later, which means that the universe expands first 
and then later shrinks.  In Fig.3, $\phi$ oscillates and $\rho$ decreases 
with small oscillation which means that an oscillating shrinking universe 
is realized. The scalar oscillating curvature goes to zero. The behavior 
in Fig.3 is very similar to that found in case of the bosonic model 
reduced from 4 dimensional model with Einstein-Hilbert action \cite{KNO} 
if we  replace the two-dimensional curvature by the four-dimensional one.

In Fig.4, $\phi$ increases monotonically in time and $\rho$ increases 
first but decreases later, i.e., the universe expands first and then 
shrinks. The scalar curvature goes to vanishing.  The behavior of 
solutions when $N=100$ as in Fig.4 is not sensitive to the value of 
$\lambda^2$, which will imply that the system would be governed by the 
naive large-$N$ structure found previously.
In the large $N$ limit, Eqs.(\ref{eqnpp2}), (\ref{req2}), (\ref{eqtp2}) 
are rewritten as
\bea
\label{eqnpp3}
0&=&{1 \over 2} \e^{2\phi}A^2+{1 \over 8}
\left( \partial_t ^2 \rho 
- (\partial_t \rho)^2 \right) \nn
&& +{1 \over 2} \left( \rho+{1 \over 2}\right)
\partial_t \phi \partial_t \phi \nn
&& -{1 \over 4}\left\{ -2 \partial_t  \rho \partial_t  \phi 
+\partial_t ^2 \phi \right\} +  t_0 \\
\label{req3}
0&=&-{1 \over 8}\partial_t^2 \rho
-{1 \over 4}\partial_t  \phi \partial_t   \phi 
+{1 \over 4}\partial_t^2 \phi \\
\label{eqtp3}
0&=&- {1 \over 2}\e^{2\phi}A^2 + \left\{
-{1 \over 2}\partial_t (\rho \partial_t  \phi)
-{1 \over 4}\partial_t^2 \rho \right\} \ .
\eea
Especially in case of $A^2=0$, we can delete $\phi$ from Eqs.(\ref{eqnpp3}), 
(\ref{req3}) and (\ref{eqtp3}) and we find 
\be
\label{rreq}
(1 + \rho )(\partial_t\rho)^2-8t_0\rho= c^2\ \ (\mbox{constant})\ .
\ee
This tells us that there is a singularity when $\rho=-1$. This singularity 
corresponds also to a singularity in the scalar curvature $R$ in (\ref{sR}) 
since
\be
\label{Rs}
R=8\e^{-2\rho}\partial_+\partial_-\rho 
\sim {2\e^2\left( -8t_0 + c^2\right) 
\over 1+\rho}
\ee
 Using a new variable and the parameter
\be
\label{newvp}
\hat\rho \equiv (1 + \rho )^{3 \over 2}\ ,\ \ \ 
E\equiv {9 \over 8}c^2\ ,
\ee
we can rewrite (\ref{rreq}) as follows
\be
\label{Eeq}
{1 \over 2}(\partial_t \hat\rho)^2 - 9t_0(\hat\rho^{2 \over 3}-1)=E\ .
\ee
Therefore we can regard the system as that of a particle with unit 
mass in the potential $- 9t_0(\hat\rho^{2 \over 3}-1)$.
When $t_0>0$, the system becomes unstable and $\hat\rho$ and $\rho$ 
increase monotonically. The $t_0>0$ case corresponds to an expanding 
universe. On the other hand, when $t_0<0$, the universe shrinks 
even if it expands for some initial period and then later the universe 
``encounters'' the curvature singularity at $\rho=-1$.

Before closing this section, we make some remarks about  the duality 
structure of $N=1$ dilatonic supergravity on bosonic background.
The duality might be useful when we consider the so-called 
``graceful exit'' problem \cite{BK} for inflationary universe.

We now consider the following simplified classical action:
\be
\label{1}
S=-{1 \over 2\pi }\int d^2 x \sqrt{-g} \left[
\e^{-2\phi}
\left( R+4(\nabla \phi)^2 \right) 
 + {1 \over 2}\e^{2\alpha\phi}
\sum_{i=1}^N(\nabla \chi_i)^2\right]\ .
\ee
Here we introduced a parameter $\alpha$ and we have put $\lambda^2=0$. 
Then the trace anomaly induced effective action \cite{trace} is given by
\be
\label{qc}
W=-{1 \over 2}\int d^2x \sqrt{-g} \left[ 
{N \over 32\pi}R{1 \over \Delta}R 
-{N \over 4\pi}\alpha^2
\nabla^\lambda \phi
\nabla_\lambda \phi {1 \over \Delta}R 
+{N \over 4\pi}\alpha\phi R \right]\ .
\ee
In the conformal gauge, the actions (\ref{1}) and (\ref{qc}) take the 
following forms:
\bea
\label{SW1}
S+W&=& -{1 \over 2\pi}\int d^2x \left[\e^{2\phi}\left\{
4\partial_+\partial_-\rho - 8\partial_+\phi \partial_-\phi 
\right\} \right. \nn
&& \left. + N \left\{-{1 \over 4}\rho\partial_+\partial_-\rho
-\alpha^2\rho\partial_+\phi\partial_-\phi 
+ \alpha\phi\partial_+\partial_-\rho\right\}\right]\ .
\eea
Here we treat $S+W$ as a classical action and we solve 
the equations of motion. If we are interested in the case 
where all the fields only depend on time $t$, we can replace 
$\partial_\pm$ by ${1 \over 2}\partial_t$.
Then the action (\ref{SW1}) becomes 
\bea
\label{SW2}
S+W&=& -{1 \over 2\pi}\int d^2x \left[\e^{2\phi}\left\{
\partial_t^2 \rho - 2(\partial_t\phi)^2 
\right\} \right. \nn
&& \left. +N \left\{-{1 \over 16}\rho\partial_t^2\rho
-{\alpha^2 \over 4}\rho(\partial_t\phi)^2 
+ {\alpha \over 4}\phi\partial_t^2\rho\right\}\right]\ .
\eea
Furthermore, we replace the conformal time $t$ by the 
cosmological time $\hat t$ defined in Eq.(\ref{ct}).
If we use a new variable $R$ instead of $\rho$ 
\be
\label{r}
r=\e^{-2\phi + 2\rho}\ , 
\ee
we obtain
\bea
\label{Rp}
S+W&=&-{1 \over 2\pi}\int d^2x\left[ \dot r \dot \phi 
+ N\e^{2\phi}\left\{{1 \over 4}{\dot r^2 \over r}
+ \left({1 \over 16} - {\alpha \over 8}\right)\dot\phi\dot r 
+ \left({1 \over 16} - {\alpha \over 4}\right)r\dot\phi^2 
\right.\right. \nn
&& \left.\left. 
-{\alpha^2 \over 4}\left({1 \over 2}\ln r + \phi\right)
\dot\phi^2 \right\}\right]\ .
\eea
Here $\dot\phi = {d\phi \over d\hat t}$ and $\dot r = 
{d r \over d\hat t}$.
The variables $\phi$ and $r$ express the classical 
duality $\phi\leftrightarrow r$.
We now consider further redefinitions of the variables.
When
\be
\label{alpha}
\alpha^2=8
\ee
defining the following new variables
\bea
\label{UV}
U&=&r + N\e^{2\phi}\left({1 \over 4}\ln r + {1 \over 32}
\right) \nn
&=&\e^{2\rho - 2\phi} 
+ N\e^{2\phi}\left({1 \over 2}(\rho - \phi) + {1 \over 32}
\right) \nn
V&=&\phi + N \e^{2\phi}r\left\{ {1 \over 32}
- {\alpha \over 8}- {1 \over 3}\ln r 
+ {1 \over 3}\left(1 -2\phi\right)\right\} \nn
&=&\phi + N \e^{2\rho}\left\{ {1 \over 32}
- {\alpha \over 8}- {2 \over 3}\rho  
+ {1 \over 3}\right\}\ ,
\eea
we obtain
\be
\label{UVS}
S+W=-{1 \over 2\pi}\int d^2x \dot U \dot V + {\cal O}(N^2)\ .
\ee
This tells that the system has a duality 
\be
\label{UVdual}
U\leftrightarrow V
\ee
when $N$ is large but $N\hbar$ is small. Note that the Planck constant 
$\hbar$ is implicitly contained in (\ref{UVS}), {\it i.e.}, $\hbar$ 
was chosen to be a unity and ${\cal O}(N^2)$ means ${\cal O}(\hbar^2 
N^2)$.  In case of the bosonic model, similar argument holds and we can 
find a ``perturbative duality'' in (\ref{UVdual}) if we change the 
coefficients and parameters a little bit, e.g. $\alpha^2={16 \over 3}$ 
instead of (\ref{alpha}).

\section{4d Reduced Supergravity  Model}

~~~We now restrict our discussion to 4D action of Einstein gravity with 
a cosmological term and matter described by $N$ minimal scalars $a_i$ 
coupled to the metric with spherical symmetry:
\be
\label{spmetric}
ds^2=g_{\mu\nu}dx^\mu dx^\nu + \e^{-2\phi}d\Omega\ .
\ee
Here the two-dimensional metric and dilaton depend only on time and radius.
The spherically reduced action reads
\be
\label{3}
S_{red}=\int d^2x \sqrt{-g}\e^{-2\phi}
\left[-{1 \over 16\pi G}
\{R + 2(\nabla  \phi)^2 -2\Lambda + 2\e^{2\phi}\} 
+ {1 \over 2}\sum_{i=1}^N(\nabla \chi_i)^2 \right]\ .
\ee
In the 4-dimensional $N=1$ supergravity model, the metric (\ref{spmetric}) 
can be realized by choosing the vierbein fields $e^a_\mu$ as follows
\bea
\label{spvier}
&& e^0_{\theta, \varphi}=e^3_{\theta, \varphi}=
e^1_{t,r}=e^2_{t,r}=0\ , \nn
&& e^1_\theta=\e^\phi\ ,\ \ 
e^2_\varphi=\sin\theta\e^\phi\ ,\ \ 
e^1_\varphi=e^2_\theta=0\ .
\eea
The expression (\ref{spvier}) is unique up to local Lorentz transformation.
The local supersymmetry transformation for the vierbein field with 
the parameter $\zeta$ and $\bar\zeta$ is given by, 
\be
\label{strnsf}
\delta e^a_\mu =i\left(\psi_\mu\sigma^a \bar\zeta
- \zeta\sigma^a \bar\psi_\mu\right)\ .
\ee
Here $\psi_\mu$ is the Rarita-Schwinger field (gravitino) and we follow 
the standard notations of ref.\cite{WB} (see also \cite{GGRS}.
If we require that the metric has the form of  
(\ref{spmetric}) after the local supersymmetry transformation, i.e. 
\bea
\label{metric}
&&\delta g_{t\theta}=\delta g_{r\theta}
=\delta g_{t\varphi}=\delta g_{r\varphi}=
\delta g_{\theta\varphi}=0 \ , \nn
&& \delta g_{\varphi\varphi}
=\sin\theta \delta g_{\theta\theta}\ ,
\eea
we obtain, up to local Lorentz transformation,
\bea
\label{solpara}
&&\zeta_1 = \bar\zeta^1\ ,\ \ \zeta_2 = \bar\zeta^2\ ,\\
\label{solgrtn}
&& \psi_{\varphi 1}=\sin\theta\psi_{\theta 1}\ ,\ \ 
\bar\psi_\varphi^1=\sin\theta\bar\psi_\theta^1 \ , \nn
&& \psi_{\varphi 2}=-\sin\theta\psi_{\theta 2}\ ,\ \ 
\bar\psi_\varphi^2=-\sin\theta\bar\psi_\theta^2 \ , 
\nn
&& \bar\psi^1_\theta=-\psi_{\theta 1}\ ,\ \ 
\bar\psi^2_\theta=-\psi_{\theta 2}\ ,\nn
&&
\psi_{r1}-\bar\psi_r^1
=-2\e^{-\phi}\e^3_r\psi_{\theta 2}\ ,\ \ 
\psi_{t1}-\bar\psi_t^1
=-2\e^{-\phi}\e^3_t\psi_{\theta 2}
\ ,\nn
&& \psi_{r2}-\bar\psi_r^2
=-2\e^{-\phi}\e^0_r\psi_{\theta 1}\ ,\ \ 
\psi_{t2}-\bar\psi_t^2
=-2\e^{-\phi}\e^0_r\psi_{\theta 1}
\ .
\eea
Eq.(\ref{solpara}) shows that the local supersymmetry of the spherically 
reduced theory is $N=1$, which should be compared with the torus 
compactified case, where the supersymmetry becomes $N=2$.
Let the independent degrees of freedom of the Rarita-Schwinger fields 
be denoted by
\bea
\label{gravitino}
&& 2\psi_r^1\equiv\psi_{r1}+\bar\psi_r^1\ ,\ \ 
2\psi_r^2\equiv\psi_{r2}+\bar\psi_r^2\ ,\nn
&& 2\psi_t^1\equiv\psi_{t1}+\bar\psi_t^1\ ,\ \ 
2\psi_t^2\equiv\psi_{t2}+\bar\psi_t^2\ ,\\ 
\label{dilatino}
&&\chi_1\equiv\psi_{\theta 1}\ ,\ \ 
\chi_2\equiv\psi_{\theta 2}\ ,
\eea
then we can regard $\psi_{t,r}$ and $\chi$ as the gravitino and 
the dilatino, respectively, in the spherically reduced theory.

We concentrate on the spherically symmetric metrics of Kantowski-Sacks 
form \cite{XVI} in (\ref{3}) where $g_{\mu\nu}=a^2(t)\eta_{\mu\nu}$. Such 
a metric describes a Universe with a $S^1\times S^2$ spatial geometry 
(for some of its properties, see \cite{XVII}). In the case of dilaton 
supergravity obtained by reducing of 4d supergravity with chiral scalar
supermatter, the action in the bosonic background is given by 
\bea
\label{bsSred}
S&=&
\int d^2x e \biggl[-{\e^{-2\phi} \over 2\pi}
\biggl\{R+2\partial_\mu\phi\partial^\mu\phi
-2\Lambda^2 + 2 \e^{2\phi} \nn
&& +{1 \over 2}\sum_{i=1}^N \left(
\partial_\mu a_i \partial^\mu a_i
+ \bar\xi_i \gamma^\mu D_\mu \xi_i\right) \biggr\} \nn
&& -{1 \over 2\pi}\left\{ {N \over 32}R{1 \over \Box}R
- {N \over 4} \partial_\mu\phi\partial^\mu\phi 
{1 \over \Box} R
+ {N \over 4}\phi R\right\}\biggr]\ .
\eea
Here, the last line term is the anomaly induced effective action due 
to matter. We obtain the following equations replacing (\ref{ddp}),
(\ref{ddr}) and (\ref{t0}),
\bea
\label{ddpred}
\partial_t^2 \phi &=& \left({5 \over 2} 
-{N \over 4}\rho\e^{2\phi}\right)
(\partial_t\phi)^2 
- \left(2 + {N \over 4}
\e^{2\phi}\right)\partial_t\rho\partial_t\phi 
+ {N \over 16} \e^{2\phi} (\partial_t\rho)^2 \nn
&& - {N \over 4}\e^{4\phi}A^2 + \lambda^2\e^{2\rho}
+\e^{2\phi+2\rho} - {N \over 2}\e^{2\phi}t_0
\\
\label{ddrred}
\partial_t^2 \rho &=& \left(1 + {N \over 8}\e^{2\phi}\right)
(\partial_t\rho)^2 
- {32 \over N}\e^{-2\phi} 
 \left(1 + {N \over 8}\e^{2\phi}\right)^2
\partial_t \rho \partial_t \phi \nn
&& + 4\left\{\left({1 \over 4}+{3 \over N}\e^{-2\phi} \right)
-\left(1 + {N \over 8}\e^{2\phi}\right)\rho\right\}
(\partial_t\phi)^2 \\
&& - 4\e^{2\phi}\left(1 + {N \over 8}\e^{2\phi}\right)A^2 
-8\left(1 + {N \over 8}\e^{2\phi}\right)t_0 
+ 2\lambda^2 \e^{2\rho}\ .\nonumber
\eea
\bea
\label{t0red}
t_0&=& -\left\{-16\e^{-2\phi} + 2N 
+ {N^2 \over 4}\e^{2\phi}\left(\rho + 1 \right)
\right\}^{-1} \nn
&& \times\Bigg[
\left\{{24 \over N}\e^{-4\phi} - (8\rho + 7)\e^{-2\phi}
- {N \over 4}\left(\rho +1 \right)
+ {N^2 \over 8}\e^{2\phi}\left(\rho + 1 \right)\rho
\right\}(\partial_t\phi)^2 \nn
&& + \left\{ - {64 \over N}\e^{-4\phi} 
+N\left(\rho + {3 \over 2} \right)
+ {N^2 \over 8}\e^{2\phi}\left(\rho + 1\right)
\right\}\partial_t\rho \partial_t\phi \nn
&& + \left\{2\e^{-2\phi} - {N \over 4} 
- {N^2 \over 32}\e^{2\phi}\left(\rho + 1\right)
\right\}(\partial_t\rho)^2 \nn
&& + \left\{ -8 + {N \over 2}\e^{2\phi} + {N^2 \over 8}\e^{4\phi}
\left(\rho + 1 \right)\right\}A^2  \nn
&& +\{ 4\e^{-2\phi} - N(\rho +1)\}
{\lambda^2 \over 2}\e^{2\rho}
-\left(4\e^{-2\phi} + {N \over 2}\right)\Bigg]
\eea

Equations (\ref{ddpred}), (\ref{ddrred}) and (\ref{t0red}) can be 
also solved numerically for several parameters with the initial 
condition $\phi=\rho={d\phi \over dt} ={d\rho \over dt}=0$ at $t=0$.
Typical graphs are given in Figs.5 and 6.  We calculated the 
four-dimensional scalar curvature corresponding to the metric 
(\ref{spmetric}) which is given by
\be
\label{4dR}
R_{4d}=-\e^{-2\rho}\left\{
2\partial^2_t\rho - 4\partial^2_t\phi 
+ 6(\partial_t \phi)^2\right\}\ .
\ee
In the 4d reduced model, the curvature singularity always seems to appear, 
at least in cases under discussion.  In Fig.5, both the dilaton field 
$\phi$ and the conformal factor $\rho$ increase monotonically in time as 
in Fig.1 in CGHS type model.  If we regard the solution as describing a
universe of the Kantowski-Sacks form with the topology $S^2\times S^1$, the 
radius of $S^2$, which is given by $\e^{-\phi}$, decreases to zero but 
the radius of $S^1$, which is given by $\e^{2\rho}$, increases. Note that  
the radius of $S^1$ corresponds to the radius of the universe when we 
regard this model as a two dimensional one.  The four-dimensional scalar 
curvature increases first and goes to minus infinity.  In Fig.6, $\phi$ 
increases monotonically and $\rho$ decreases at first but increases 
infinitely after that.  This means $S^2$ shrinks but the $S^1$ factor 
shrinks at first and later expands.  The four-dimensional scalar curvature 
goes to zero at first but increases to infinity later.  The dilaton 
field $\phi$ always increases and runs away to the singularity, 
which means  $S_2$ in the universe of Kantowski-Sacks form shrinks 
to a point, which causes the curvature singularity. When we regard 
the model as a two-dimensional one, the radius of the universe 
goes to infinity at the final stage in the cases studied here.
We see that at $t=0$ there is no singularity in most of our cases.
However, at late times there is as a rule singularity.
Note that the behaviour of our cosmologies at late times is not so important 
as other effects should define late time structure of the Universe.

\section{$N=2$ Dilaton Supergravity Model}

~~~In this section, we discuss $N=2$ dilaton supergravity. Notations 
and conventions are mainly following to \cite{GATES}. However, the 
possibility of considering a purely twisted matter potential term 
point out in \cite{GATES} is not treated. The general classical 
action of $U_A(1)$ $N=2$ dilaton supergravity with matter is given 
by\footnote{We use $N=1$ or $N=2$ to denote type of SG but also we use 
$N$ as number of matter multiplets which should not cause confusion in 
the context used.}
\bea
\label{cS}
S&=&{1 \over 2\pi}\int d^2x d^2\theta \cE C(\Phi) R 
+ {1 \over 2} \int d^2x d^4\theta^2 
E^{-1} Z(\Phi) Z(\bar\Phi) \nn
&& + \int d^2x d^2\theta \cE V(\Phi) 
+ {1 \over 2}\int d^2x d^4\theta
E^{-1} g(\Phi) \sum_{i=1}^M \Sigma_i \bar\Sigma_i \nn
&& +\int d^2x d^4\theta E^{-1} f(\Phi)\sum_{i=1}^N 
\bar{\cal X}_i{\cal X}_i 
+ h.c. \nn 
&=& {1 \over 2\pi}\int d^2x e^{-1} \Biggl[ C'(\phi)FB 
+ C'(\bar\phi)\bar F\bar B  \nn
&& + 2(C(\phi)+C(\bar\phi))(B\bar B - \cR )
+ 2i(C(\phi)-C(\bar\phi))\cF \nn
&& +Z'(\phi)Z'(\bar\phi)(- F\bar F 
- \partial_+ \phi \partial_- \bar\phi)
+ (V'(\phi)F + V'(\bar\phi)\bar F \nn
&& - \sum_{i=1}^M \left\{
(g'(\phi)F a_i + g(\phi) G_i )\bar G_i
+G_i ( g'(\bar\phi)\bar F \bar a_i + g(\bar\phi)\bar G_i) 
\right. \nn
&& \left. +\partial_+ (g(\phi)a_i)\partial_- \bar a_i
+\partial_+ a_i \partial_- (g(\bar\phi)\bar a_i)\right\} \nn
&& -(f(\phi) + f(\bar\phi))\sum_{i=1}^N \left\{\bar H_i H_i 
- {1 \over 2}\left( \partial_-\bar \chi_i 
\partial_+\chi_i 
+\partial_+\bar \chi_i \partial_-\chi_i \right)\right\}
\Biggr] \nn
&& + \mbox{fermionic terms}
\eea
Here $\Phi=(\phi, \chi, F)$, $\Sigma_i=(a_i, \xi_i, G_i)$ 
and ${\cal X}=(\chi_i, \zeta_i, H_i)$ 
are the dilaton, chiral matter and twisted chiral matter 
multiplets, respectively, 
$\cR$ is the scalar curvature
and $\cF=\partial_0 A_1 - \partial_1 A_0$ is the field strength of the 
vector field $A_\mu$. The action (\ref{cS}) shows that the imaginary 
part of $C(\phi)$ can be identified with the axion field. Integrating 
out $A_\mu$, we find
\be
\label{im}
C(\phi)-C(\bar\phi)=c\ \ (\mbox{constant})
\ee
If we consider the solution where $c=0$, the dilaton field becomes real
\be
\label{dr}
\phi=\bar\phi\ .
\ee
If we consider the case of $N\gg M$, $1$, the bosonic part of the 
effective action would be given by \cite{trace} 
\be
\label{qc2}
W=-{1 \over 2}\int d^2x \sqrt{g} \Bigl[ 
{N \over 16\pi}\left(
R{1 \over \Box}R - \cF{1 \over \Box}\cF\right)
-{N \over 2\pi}\nabla^\lambda \varphi
\nabla_\lambda \varphi {1 \over \Box}R 
+{N \over 2\pi}\varphi R \Bigr]\ .
\ee
Here
\be
\label{varphi}
\varphi\equiv -{1 \over 2}\ln \left( {f(\phi)+f(\bar\phi) \over 2}
\right)
\ee
and we used that the action of the fermion component of ${\cal X}_i$ 
in the bosonic background (\ref{cS}) has the following form
\bea
\label{zetaction}
S_\zeta&=&i\int d^2x e^{-1} (f(\phi) + f(\bar\phi))\sum_{i=1}^N 
\biggl\{(\partial_\pp\zeta_-)\zeta_{\dot -}
+(\partial_=\zeta_+)\zeta_{\dot +} \nn
&& +{i \over 2}(A_\pp\zeta_-\zeta_{\dot -} - A_=\zeta_+\zeta_{\dot +})
\biggr\}
\eea
The action (\ref{qc2}) is obtained by replacing $N$ by $2N$ in the 
$N=1$ supergravity effective action and adding the second term coming 
from chiral anomaly. This is because the dilaton dependent function 
$(f(\phi) + f(\bar\phi))$ in front of the matter fermion kinetic term in 
Eq.(\ref{zetaction}) can be absorbed into the redefinition of $\zeta$ 
and the matter fermions do not give any contribution to the dilaton 
dependent terms of effective action.  The contribution of the energy 
momentum tensor from the matter $\zeta_i$ fermions is, as in $N=1$ case, 
absorbed into the redefinition of $t_0$.  Therefore the behavior of the 
cosmological solution is almost identical with the $N=1$ case in the 
bosonic background with $\cF=0$.

We now consider the case where $\cF\neq 0$. For this case, we parametrize 
the vector field $A_\mu$ as follows;
\be
\label{theta}
A_\mu=\partial_\mu \tilde\theta 
+{16 \over N}g_{\mu\nu}\epsilon^{\nu\rho}\partial_\rho \theta\ .
\ee
Note that $\tilde\theta$ expresses the gauge degree of freedom.  Then we 
can rewrite the chiral anomaly term in Eq.(\ref{qc2}) in a local form,
\be
\label{lchianmly}
{N \over 32\pi}\int d^2x \sqrt{g}  
\cF{1 \over \Delta}\cF
={8 \over N\pi}\int d^2x \sqrt{g}  \theta\Box\theta\ .
\ee
The $\theta$-equation of motion obtained from $S+W$ has the following form 
\be
\label{theq}
0=\Box\left\{
2i(C(\phi)-C(\bar\phi))+\theta \right\}
\ee
that is, 
\be
\label{theq2}
\theta = - 2i(C(\phi)-C(\bar\phi)) + 
\mbox{constant}\ .
\ee
Now we consider the case,
\be
\label{cghstype}
C(\phi)=-{1 \over 4}\e^{-2\phi},\ \ \ 
Z(\phi)=4\e^{-\phi}, \ \ \ 
V(\phi)=\Lambda \e^{-2\phi},\ \ \ 
f(\phi)=\e^{-2\phi}
\ee
and the background where the chiral matter multiplet $\Sigma_i$ vanishes.
Then integrating the auxilliary fields $B$, $\bar B$, $F$ and $\bar F$ 
and using (\ref{theq2}), we can rewrite the bosonic part of $S+W$ as follows:
\bea
\label{SpW}
&& S+W \nn
&=&{1 \over 2\pi}\int d^2x e^{-1} \Biggl[
-\e^{-2\varphi}\cR - {1 \over \sqrt{\e^{-4\varphi} + \theta^2}}
\left\{4\e^{-4\varphi}\partial_\mu\varphi \partial^\mu \varphi
+ \partial_\mu\theta \partial^\mu \theta\right\} \nn
&& + {4\Lambda^2 ( \e^{-4\varphi} + \theta^2 ) \over 
-{1 \over 4}(\e^{-4\varphi} + \theta^2)\e^{2\varphi}
+ 16 \sqrt{\e^{-4\varphi} + \theta^2}} 
- {16 \over N}  \theta\Box\theta 
+2\e^{-2\varphi} \sum_{i=1}^N\partial_\mu\bar\chi_i\partial^\mu\chi_i
\Biggr] 
\nn
&& -{1 \over 2}\int d^2x \sqrt{g} \Bigl[ 
{N \over 16\pi}R{1 \over \Box}R 
-{N \over 2\pi}\nabla^\lambda \varphi
\nabla_\lambda \varphi {1 \over \Box}R 
+{N \over 2\pi}\varphi R \Bigr]\ .
\eea
Here $\theta$ should be understood  not to represent the degrees of 
freedom of the vector field but to be the imaginary part of $\e^{-2\phi}$  
using ({\ref{theq2}) {\it i.e.} to be the axion field.
Then the equations corresponding to (\ref{eqnpp2}) -- (\ref{mat2})
become ($T$ is absorbed in the redefinition of $t_0$)
\bea
\label{eqnppn2}
0&=&4\e^{-2\varphi}\partial_t \rho \partial_t \varphi 
- {1 \over 2\sqrt{\e^{-4\varphi} + \theta^2 }}\left( 
4\e^{-4\varphi}\left(\partial_t \varphi\right)^2 
+\left(\partial_t\theta\right)^2\right) 
- {4 \over N}\left(\partial_t\theta\right)^2 \nn
&& + {1 \over 2} \e^{-2\varphi}\sum_{i=1}^N 
\partial_t  \chi_i \partial_t  \bar\chi_i  
 +{N \over 4}\left( \partial_t ^2 \rho 
- (\partial_t \rho)^2 \right) + N 
\left( \rho+{1 \over 2}\right)
(\partial_t \varphi)^2 \nn
&& -{N \over 2}
\left\{ -2 \partial_t  \rho \partial_t  \varphi 
+\partial_t ^2 \varphi \right\} + 2N t_0 \\
\label{reqn2}
0&=&\e^{-2\varphi}\left(2\partial_t^2\varphi 
-4 (\partial_t\varphi)^2 \right) 
-{4\Lambda^2 ( \e^{-4\varphi} + \theta^2 )\e^{2\rho} 
\over -{(\e^{-4\varphi} + \theta^2)\e^{2\varphi} \over 4}
+ 16 (\e^{-4\varphi} + \theta^2)^{1 \over 2}} \nn
&& -{N \over 4}\partial_t^2 \rho
-{N \over 2}(\partial_t  \varphi)^2  
+{N \over 2}\partial_t^2 \varphi \\
\label{eqtpn2}
0&=& -{4 ( \e^{-4\varphi} + 2\theta^2 )\e^{-4\varphi} 
\over (\e^{-4\varphi} + \theta^2 )^{3 \over 2}} (\partial_t\varphi)^2
+ {4\e^{-4\varphi} \over 
( \e^{-4\varphi} + \theta^2 )^{1 \over 2}}
\partial_t^2\varphi \nn
&& -{\e^{-4\varphi} \over ( \e^{-4\varphi} + \theta^2 )^{3 \over 2}}
(\partial_t\theta)^2 
-{4\theta\e^{-4\varphi} 
\over ( \e^{-4\varphi} + \theta^2 )^{3 \over 2}}
\partial_t\theta\partial_t\varphi 
+2\e^{-2\varphi}\partial_t^2  \rho \nn
&& -2\Lambda^2 \e^{2\rho} 
\left\{-{(\e^{-4\varphi} + \theta^2)\e^{2\varphi} \over 4}
+ 16 (\e^{-4\varphi} + \theta^2)^{1 \over 2}\right\}^{-2} \nn
&& \times\left\{{1 \over 2}(\e^{-4\varphi} + \theta^2)^2
\e^{2\varphi}-32(\e^{-4\varphi} 
+ \theta^2)^{1 \over 2}\e^{-4\varphi}\right\} \nn
&& - {1 \over 2}\e^{-2\varphi}
\sum_{i=1}^N  \partial_t  \chi_i \partial_t  \bar\chi_i  
+ \left\{
-N\partial_t (\rho \partial_t  \varphi)
-{N \over 2}\partial_t^2 \rho \right\} \\
\label{theqn2}
0&=& {\theta \over (\e^{-4\varphi} + \theta^2)^{3 \over 2}}
(\partial_t\theta)^2
+ \left\{ {4 \over (\e^{-4\varphi} + \theta^2)^{1 \over 2}}
- {8 \over N}\right\}\partial_t^2\theta \nn
&& + {2\theta\e^{-4\varphi}  \over (\e^{-4\varphi} 
+ \theta^2)^{3 \over 2}}
(\partial_t\varphi)^2 + {2\e^{-4\varphi}  
\over (\e^{-4\varphi} + \theta^2)^{3 \over 2}}
\partial_t\varphi \partial_t\theta \nn
&& - 32\Lambda^2\e^{2\rho}\theta (\e^{-4\varphi} + \theta^2)^{1 \over 2}
\left\{-{(\e^{-4\varphi} + \theta^2)\e^{2\varphi} \over 4}
+ 16 (\e^{-4\varphi} + \theta^2)^{1 \over 2}\right\}^{-2} \\
\label{matn2}
0&=& \partial_t (\e^{-2\varphi}\partial_t \chi_i) 
=\partial_t (\e^{-2\varphi}\partial_t \bar\chi_i)\ .
\eea
Eq.(\ref{matn2}) can be also integrated to be 
$\partial_t \chi_i=A_i\e^{2\varphi}$ 
($\partial_t \bar\chi_i=\bar A_i\e^{2\varphi}$).
 Defining $A^2={1 \over 2N}\sum_{i=1}^N A_i \bar A_i$, Eqs.(\ref{eqnppn2}), 
(\ref{reqn2}), (\ref{eqtpn2}) and (\ref{theqn2}) are rewritten as 
follows,
\bea
\label{vpn2}
\partial_t^2\varphi &=& -2
\left(1+ {N \over 4}\e^{2\varphi}\right)
\partial_t\rho\partial_t\varphi
+\left({\e^{-2\varphi} \over (\e^{-4\varphi} + \theta^2)^{1 \over 2}}
+2 - {N \over 2}\rho\e^{2\varphi}\right) 
(\partial_t\varphi)^2 \nn
&& + \left( {1 \over 4 (\e^{-4\varphi} 
+ \theta^2)^{1 \over 2}}
+ {2 \over N}\right)\e^{2\varphi}
(\partial_t\theta)^2
+{N \over 8}\e^{2\varphi}(\partial_t\rho)^2 \nn
&& -{N \over 2}\e^{4\varphi}A^2
+{2\Lambda^2(\e^{-4\varphi} + \theta^2)\e^{2\rho + 2\varphi} 
\over -{(\e^{-4\varphi} + \theta^2)\e^{2\varphi} \over 4}
+16(\e^{-4\varphi} + \theta^2)^{1 \over 2}}
- N t_0\e^{2\varphi} \\
\label{rn2}
\partial_t^2 \rho &=& -{16 \over N}\left(1+ {N \over 4}
\e^{2\varphi}\right)^2\e^{-2\varphi}\partial_t\rho
\partial_t\varphi \nn
&& + \left\{ 2+{8 \over N}\e^{-2\varphi}
\left(1+ {N \over 4}\e^{2\varphi}\right)\left(
{\e^{-2\varphi} \over (\e^{-4\varphi} + \theta^2)^{1 \over 2}}
- {N \over 2}\rho\e^{2\varphi}\right)\right\}
(\partial_t\varphi)^2 \nn
&& + {8 \over N}\left(1+ {N \over 4}\e^{2\varphi}\right)
\left({1 \over 4(\e^{-4\varphi} + \theta^2)^{1 \over 2}}
+ {2 \over N}\right)(\partial_t\theta)^2 \nn
&& +\left(1+ {N \over 4}\e^{2\varphi}\right)(\partial_t\rho)^2
-4\e^{2\varphi}\left(1
+ {N \over 4}\e^{2\varphi}\right)A^2 \nn
&& +{ 4\Lambda^2\e^{2\rho+2\varphi}
(\e^{-4\varphi} + \theta^2)
\over - { (\e^{-4\varphi} + \theta^2)\e^{2\varphi} 
\over 4} + 16 (\e^{-4\varphi} + \theta^2)^{1 \over 2}} \nn
&& - 8 \left(1+ {N \over 4}\e^{2\varphi}\right)t_0 \\
\label{thn2}
\partial_t^2\theta &=& \left\{ -{4 \over (\e^{-4\varphi} 
+ \theta^2)^{1 \over 2}}+ {8 \over N}\right\}^{-1}\Biggl[
{\theta \over (\e^{-4\varphi} + \theta^2)^{3 \over 2}}
(\partial_t\theta)^2
\nn
&& + {2\theta\e^{-4\varphi}  \over (\e^{-4\varphi} 
+ \theta^2)^{3 \over 2}}
(\partial_t\varphi)^2 + {2\e^{-4\varphi}  
\over (\e^{-4\varphi} + \theta^2)^{3 \over 2}}
\partial_t\varphi \partial_t\theta \\
&& - 32\Lambda^2\e^{2\rho}\theta (\e^{-4\varphi} + \theta^2)^{1 \over 2}
\left\{-{(\e^{-4\varphi} + \theta^2)\e^{2\varphi} \over 4}
+ 16 (\e^{-4\varphi} + \theta^2)^{1 \over 2}\right\}^{-2} \Biggr] \nn
\label{t0n2}
t_0&=&\left\{{4N\e^{-2\varphi} \over 
(\e^{-4\varphi} + \theta^2)^{1 \over 2}} 
+ 16\e^{-2\varphi}\left(1 - {N^2 \over 16}\e^{4\varphi}
\right) - N^2\e^{2\varphi}\rho \right\}^{-1} \nn
&& \times\Biggl[ \Biggl\{-{4 (\e^{-4\varphi} + 2\theta^2) 
\e^{-4\varphi} \over (\e^{-4\varphi} + \theta^2)^{3 \over 2}}
+ { 4\e^{-6\varphi} \over \e^{-4\varphi} + \theta^2}
+4\e^{-2\varphi} - N \nn
&& +{8\e^{-4\varphi}+ {16 \over N}\e^{-6\varphi}
\left(1 - {N^2 \over 16}\e^{4\varphi}\right) 
\over (\e^{-4\varphi} + \theta^2)^{1 \over 2}} \nn
&& +\left\{- {3N\e^{-2\varphi} 
\over (\e^{-4\varphi} + \theta^2)^{1 \over 2}} 
-2N - 8\e^{-2\varphi}
\left(1 - {N^2 \over 16}\e^{4\varphi}\right) \right\}\rho \nn
&& + {N^2 \over 2}\e^{2\varphi}\rho^2\Biggr\}
(\partial_t\varphi)^2 \nn
&& + \Biggl\{ - {8\e^{-4\varphi} \over 
(\e^{-4\varphi} + \theta^2)^{1 \over 2}} 
\left(1
+ {N \over 4}\e^{2\varphi}\right) - N \nn
&& - {32 \over N}\e^{-4\varphi}
\left(1 - {N \over 4}\e^{2\varphi}\right)
\left(1 + {N \over 4}\e^{2\varphi}\right)^2 \nn
&& + 2N \left(1 + {N \over 4}\e^{2\varphi}\right)\rho 
\Biggr\}\partial_t\rho \partial_t\varphi \nn
&& + \left\{ {N\e^{-2\varphi} \over 2 
(\e^{-4\varphi} + \theta^2)^{1 \over 2}} 
+2 \e^{-2\varphi} \left(1 - {N^2 \over 16}\e^{4\varphi}\right)
- {N^2 \over 8}\e^{2\varphi}\rho \right\}
(\partial_t\rho)^2 \nn
&& +\Biggl\{ - {\e^{-4\varphi} \over  
(\e^{-4\varphi} + \theta^2)^{3 \over 2}}
+ {\e^{-2\varphi} \over (\e^{-4\varphi} + \theta^2)}
+ {{12 \over N}\e^{-2\varphi} 
- {N \over 4} \e^{2\varphi} \over 
(\e^{-4\varphi} + \theta^2)^{1 \over 2}} \nn
&& + {32 \over N^2} 
\e^{-2\varphi} \left(1 - {N^2 \over 16}\e^{4\varphi}\right)
- N \e^{2\varphi}\left({1 \over  
4(\e^{-4\varphi} + \theta^2)^{1 \over 2}}
+ {2 \over N}\right)\rho\Biggr\}(\partial_t\theta)^2 \nn
&& -{4\theta\e^{-4\varphi} \over  
(\e^{-4\varphi} + \theta^2)^{3 \over 2}}
\partial_t\theta\partial_t\varphi \nn
&& -{2\Lambda^2\e^{2\rho} 
\left( { (\e^{-4\varphi} + \theta^2)^2\e^{2\varphi} 
\over 2} -32 (\e^{-4\varphi} + \theta^2)^{1 \over 2}
\e^{-4\varphi} \right) \over \left(
- { (\e^{-4\varphi} + \theta^2)\e^{2\varphi} 
\over 4} + 16 (\e^{-4\varphi} + \theta^2)^{1 \over 2}
\right)^2} \nn
&& + {2\Lambda^2\e^{2\rho+2\varphi} (\e^{-4\varphi} 
+ \theta^2)
\left({4 \e^{-4\varphi}
\over (\e^{-4\varphi} + \theta^2)^{1 \over 2}}
- N \rho + 4 \e^{-2\varphi} - N \right)
\over
- { (\e^{-4\varphi} + \theta^2)\e^{2\varphi} 
\over 4} + 16 (\e^{-4\varphi} + \theta^2)^{1 \over 2}} \\
&& + \left\{ -{ 2N \over (\e^{-4\varphi} + \theta^2)^{1 \over 2}}
+ {N^2 \over 2}\e^{4\varphi} \rho
- 8 \left(1 - {N^2 \over 16}\e^{4\varphi}\right)
- N\e^{2\varphi} \right\}A^2 \Biggr]\ . \nonumber
\eea
Equations (\ref{vpn2}), (\ref{rn2}), (\ref{thn2}) and (\ref{t0n2}) can 
be also solved numerically for several parameters with the initial 
condition $\varphi=\rho={d\varphi \over dt}={d\rho \over dt}=0$, $\theta=1$ 
and ${d\theta \over dt}=1.1$ at $t=0$.  Typical graphs are given in 
Figs.7 -- 10.   In most of cases, $\theta$ increases linearly with 
respect to the conformal time $t$.  As in the 4d reduced model of Figs.5 
and 6, the singularity always seems to appear, at least in the cases 
under discussion. As a specific feature in the $N=2$ model, there 
is a case (Fig.10) where the dilaton field $\varphi$ goes to minus 
infinity. Such a case has not been found in the $N=1$ model of Figs.1 -- 6 
but was found in the bosonic model \cite{KNO}. When $\lambda^2<0$ in the 
bosonic model, the dilaton field $\varphi$ goes to minus infinity but 
note that $\lambda^2$ or $\Lambda^2$ is always positive or zero in the 
supersymmetric model. If we regard this $N=2$ model as a four dimensional 
model with the topology $S^2\times S^1$ of the Kantowski-Sacks form, the 
radius of $S^2$, which would be given by $\e^{-\varphi}$, goes to infinity.

In Fig.7, the dilaton field $\varphi$ increases monotonically in time and 
goes to infinity, the conformal factor $\rho$ decreases at first but 
increases after that and goes to infinity and $\theta$ increases linearly 
in the conformal time $t$. In Fig.8, $\varphi$ increases monotonically 
in time and goes to infinity, $\rho$ increases at first but decreases 
after that and goes to minus infinity and $\theta$ increases linearly 
in the conformal time $t$. Fig.9 is the case that $\theta$ increases 
monotonically in time but not linearly with respect to the conformal 
time $t$. In Fig.9, both of $\varphi$ and $\rho$ decrease at first and 
increase after that and go to infinity. This case should also be compared 
with that of the $N=1$ models, in most these the dilaton field $\varphi$ 
increases monotonically in time except in the case of Fig.3. Even in Fig.3, 
$\varphi$ cannot be negative.  In Fig.10, the dilaton field $\varphi$ 
oscilates a little and finally goes to minus infinity, the conformal 
factor $\rho$ decreases monotonically and goes to minus infinity and 
$\theta$ increases linearly in the conformal time $t$. In the cases of 
Figs.7 and 9, the radius of the two dimensional universe, which is given 
by $\e^\rho$, goes to infinity, that is, the universe expands. On the 
other hand, in the cases of Figs.8 and 10, the two dimensional universe 
shrinks to a point. It is, however, not clear what determines the 
fate of the two dimensional universe since it seems that there is 
no any clear correlation between the behavior of $\rho$ and 
the ``cosmological constant'' $\Lambda^2$ and/or the matter 
energy density $A^2$.
We again see that at $t=0$ there is no singularity in most of our cases.
However,  singularity appears in course of evolution. We again may argue 
that presented effects are getting negligible at late times where 
quantum corrections almost play no role.


\section{String inspired model in 4 dimensions}

~~~The low-energy effective action of the dilaton-gravity sector in the 
 superstring would be given by
\be
\label{dgaction}
S=\int d^4x \sqrt{-g} \e^{-\phi-\phi^*} \left(R-6\nabla_\mu\phi
\nabla^\mu\phi^*\right) + \cdots\ .
\ee
Here $\cdots$ denotes the terms containing moduli, gauge fields, which depend 
on the detail of the compactification, and fermionic fields.  A special
4d dilatonic supergravity coming from string theories has been considered 
in \cite{GMOV} (recently it has been independently rediscovered in 
\cite{BSBGW}). This model specifically (and is expected to be most directly 
related to superstring and heterotic strings) uses a real linear multiplet 
instead of the more familiar chiral multiplet description that follows.
We now consider the system where the massless matter fields couple
with the dilaton gravity system (\ref{dgaction}).

We first consider the case that the matter does not 
couple with the dilaton. 
In this case, the action of the matter field (WZ action) is given 
by
\be
\label{L1}
{\cal L}=\int d^2 \Theta^2 2{\cal E}\left[
-{1 \over 8}(\bar{\cal D}\bar{\cal D}-8R)\Phi^+\Phi
\right] + \mbox{h.c.}\ .
\ee
Here we used the notations of \cite{WB}.
Then the bosonic part of the action is given by
\bea
\label{L2}
{\cal L}^b&=&{1 \over 6}e AA^* {\cal R}
- e \partial_m A \partial^m A \nn
&& + {1 \over 9}(MA^* - 3 F^*)(M^*A - 3 F)
-{1 \over 9}e AA^* b_a b^a \nn
&& - {i \over 3}(A^*\partial_m A - A \partial_m A^*)b^m \ .
\eea
Note that the coefficient ${1 \over 6}$ in the first term, 
which assures the conformal invariance.
The action (\ref{L2}) is, in fact, invariant under the 
following (local) conformal transfomation with the 
parameter $\sigma$:
\bea
\label{cnftrsf}
g_{mn} &\rightarrow& g_{mn} \e^{2\sigma} \nn
(e^a_m&\rightarrow& e^a_m \e^\sigma )\nn
M &\rightarrow& M \e^{-\sigma} \nn
b^m &\rightarrow & b^m \e^{-\sigma} \nn
A & \rightarrow& A \e^{-\sigma} \nn
F & \rightarrow& F \e^{-2\sigma}\ .
\eea
The terms containing fermion fields are given by
\be
\label{L3}
{\cal L}^f=-{i \over 2}e[\chi\sigma^m{\cal D}_m\bar\chi 
+\bar\chi\bar\sigma^m{\cal D}_m\chi] + 
\mbox{terms containing gravitino}\ .
\ee
Here
\bea
\label{cov}
{\cal D}_m\bar\chi&=&\left(\partial_m 
- \bar\sigma^{nl}\omega_{mnl}\right)\bar\chi \nn
{\cal D}_m\chi&=&\left(\partial_m 
- \sigma^{nl}\omega_{mnl}\right)\chi \nn
\omega_{mnl}&=&{1 \over 2}\biggl\{
-e_{la}(\partial_n e_m^a -\partial_m e_n^a) 
-e_{ma}(\partial_l e_n^a -\partial_n e_l^a) \nn
&& +e_{na}(\partial_m e_l^a -\partial_l e_m^a) \biggr\} \nn
\bar\sigma^{mn}&=&{1 \over 4}\left(\bar\sigma^m\sigma^n 
- \bar\sigma^n\sigma^m\right) \nn
\sigma^{mn}&=&{1 \over 4}\left(\sigma^m\bar\sigma^n 
- \sigma^n\bar\sigma^m\right) 
\eea
Under the conformal transformation (\ref{cnftrsf}),
$\omega_{mnl}$ transform as
\be
\label{omega}
\omega_{nml}\rightarrow \e^{2\sigma}
\left(\omega_{mnl} + g_{nl}\partial_m \sigma
- g_{nm}\partial_l\sigma \right)\ .
\ee
Then under the conformal transformation (\ref{cnftrsf}) and 
\be
\label{chitr}
\chi \rightarrow \chi \e^{-{3 \over 2}\sigma}\ ,
\ee
the action (\ref{L3}) transform as 
\bea
\label{Lftrsf}
{\cal L}^f&\rightarrow& {\cal L}^f + \delta {\cal L}^f \nn
 \delta{\cal L}^f&\propto& (\chi\sigma^m\bar\chi
+ \bar\chi\bar\sigma^m\chi)\partial_m\sigma\ .
\eea
Since $\chi\sigma^n\bar\psi= - \psi\bar\sigma^n\chi$, $\delta{\cal L}^f$ 
vanishes and ${\cal L}^f$ is invariant under the conformal transformation.

The conformal invariance holds if we introduce dilaton (chiral) multiplet 
$\Psi$ whose first component $\phi$ is invariant under the conformal 
transformation (\ref{cnftrsf}) and the last component $G$ is transformed 
as
\bea
\label{phiG}
\phi&\rightarrow& \phi \nn
G&\rightarrow & G\e^{-\phi}\ .
\eea
as follows
\bea
\label{L4}
{\cal L}&=&\int d^2 \Theta^2 2{\cal E}\left[
-{1 \over 16}(\bar{\cal D}\bar{\cal D}-8R)
(\Phi^+(f(\Psi)\Phi)+ (f(\Psi^+)\Phi^+)\Phi)\right]
\nn
&=&{1 \over 12}e \left(f(\phi)+f(\phi^*)\right)AA^* {\cal R}\nn
&& - {1 \over 2}e \left(\partial_m (f(\phi) A) \partial^m A^* 
+ \partial_m A \partial^m (f(\phi^*)A^*) \right)\nn
&& -{i \over 4}e(f(\phi)+f(\phi^*))
[\chi\sigma^m{\cal D}_m\bar\chi
+\bar\chi\bar\sigma^m{\cal D}_m\chi ] \nn
&& + {1 \over 18}\{(MA^* - 3 f(\phi^*)F^*-3 f'(\phi^*)G^*A^*)
(M^*A - 3 F) \nn
&& + (MA^* - 3 F^*)
(M^*A - 3 f(\phi)F-3 f'(\phi)GA)\} \nn
&& -{1 \over 18}e \left(f(\phi)+f(\phi^*)
\right)AA^* b_a b^a \nn
&& - {i \over 6}
\{(A^*\partial_m (f(\phi)A) - f(\phi)A \partial_m A^*)b^m \nn
&& + (f(\phi^*)A^*\partial_m A 
- A \partial_m (f(\phi^*)A^*))b^m\} \ .
\eea
Here we put the fermionic component of $\Psi$ to vanish since we 
now regard $\Psi$ as a background.

If we consider the background where the dilaton field $\phi$ is real, 
the non-local effective action $W$ induced by the conformal anomaly 
of the dilaton coupled matter fields has been calculated in 
ref.\cite{2d4d} and is given by:
\bea
\label{v10}
W&=&b\int d^4x \sqrt{-g} F\sigma \nn
&& +b'\int d^4x \sqrt{-g} \Bigl\{\sigma\left[
2\Box^2 + 4 R^{\mu\nu}\nabla_\mu\nabla_\nu 
- {4 \over 3}R\Box + {2 \over 3}(\nabla^\mu R)\nabla_\mu 
\right]\sigma \nn
&& + \left(G-{2 \over 3}\Box R\right)\sigma \Bigr\} \nn
&& -{1 \over 12}\left(b'' + {2 \over 3}(b + b')\right)
\int d^4x \sqrt{-g}\left[R - 6 \Box \sigma 
- 6(\nabla \sigma)(\nabla \sigma) \right]^2 \nn
&& + \int d^4x \sqrt{-g} \Bigl\{ 
{a_1 \over 8} 
[(\nabla \varphi) (\nabla \varphi)]^2 \sigma 
+{a_2 \over 4}\Box \left((\nabla \varphi) (\nabla \varphi) 
\right)\sigma \nn
&& \hskip 2cm 
+ {a_2 \over 4} (\nabla \varphi) (\nabla \varphi) [(\nabla \sigma) 
(\nabla \sigma )]\Bigr\}
\eea
Here the $\sigma$-independent terms are dropped, the last terms represent 
the contribution from the dilaton dependent terms in trace anomaly. The 
dilaton dependent function $\varphi$ is defined by
\be
\label{vphi}
\varphi
=-{1 \over 2}\ln \left( {f(\phi) + f(\phi^*) \over 2} \right)
\ee
The coefficients are given by
\bea
\label{v8}
b={ 1 \over 120 (4\pi)^2 }\times 4N\ ,\ \ 
b'=-{ 1 \over 360 (4\pi)^2 }\times {13N \over2}\ ,\ \nn 
a_1={ 1 \over 32 (4\pi)^2 }\times 2N\ ,\ \ 
a_2={ 1 \over 24 (4\pi)^2 }\times 2N\ .
\eea
The coefficient $b''$  can be changed by a finite renormalization of local 
counterterm and we can put $b''=0$ (as is obtained also 
from direct calculation).  We consider below $N$ WZ multiplets 
and apply the large $N$ expansion which is why all the coefficients above 
have the multiplier $N$. In the following, we only consider the case where 
$f(\phi)$ is given by
\be
\label{fphi}
f(\phi)=\e^{-2\phi}
\ee
and we assume $\phi$ is real and that the conformally flat fiducial metric 
is $g_{\mu\nu}=\e^{2\sigma}\eta_{\mu\nu}$. Then the total action has the 
following form:
\bea
\label{v13} 
S&=&W + S_{cl} \nn
&=&\int d^4x \Bigl\{
2b'(\Box \sigma )^2 
-\left[3b'' + 2(b + b')\right]\left[
\Box \sigma + (\partial_\mu \sigma)^2\right]^2 \nn
&& +{a_1 \over 8}
[\partial_\mu \varphi \partial^\mu \varphi]^2 \sigma 
+ {a_2 \over 4}
\Box \left(\partial_\mu \varphi \partial^\mu \varphi) \right) 
\sigma \nn
&& + {a_2 \over 4}
\partial_\mu \varphi \partial^\mu\varphi 
\partial_\nu \sigma \partial^\nu \sigma \nn
&& - 6 \e^{4\sigma-2\varphi}\partial_\mu\varphi 
\partial^\mu \varphi 
+ \e^{2\sigma-2\varphi} 
\left[
-6\Box\sigma - 6 (\partial_\mu\sigma )
(\partial^\mu\sigma)\right] \Bigr\}\ .
\eea
The equations of the motion given by the variation of $\sigma$ and $\phi$ are 
given by
\bea
\label{sigmaeq}
0&=& 4b'\Box^2 \sigma \nn 
&& - 2[3b'' + 2 (b + b')]\left[
\Box \left[
\Box \sigma + (\partial_\mu \sigma)^2\right]
-2\partial_\mu\left\{\partial^\mu \sigma 
\left[\Box \sigma + (\partial_\mu \sigma)^2\right]\right\}\right] \nn
&& +{a_1 \over 8}
[\partial_\mu \varphi \partial^\mu \varphi]^2  
+ {a_2 \over 4}
\Box \left(\partial_\mu \varphi \partial^\mu \varphi \right)  \nn
&& - {a_2 \over 2}
\partial_\nu(\partial^\nu\sigma\partial_\mu \varphi 
\partial^\mu\varphi ) \nn
&& - 24 \e^{4\sigma-2\varphi}\partial_\mu\varphi 
\partial^\mu \varphi 
+ 2\e^{2\sigma-2\varphi} 
\left[
-6\Box\sigma - 6 (\partial_\mu\sigma )
(\partial^\mu\sigma)\right] \nn
&& -6 \Box \e^{2\sigma-2\varphi} 
+ 12 \partial_\mu (\e^{2\sigma-2\varphi}
\partial^\mu\sigma) \ . \\
\label{vpeq}
0&=& {a_1 \over 2}\partial_\nu \{\partial^\nu\varphi
[\partial_\mu \varphi \partial^\mu \varphi]\sigma\} 
- {a_2 \over 2}
\partial_\nu(\partial^\nu\varphi \Box \sigma ) \nn
&& - {a_2 \over 2}
\partial_\mu (\partial^\mu\varphi 
\nabla_\nu \sigma \nabla^\nu \sigma) \nn
&& + 12 \partial_\mu ( \e^{4\sigma -2\varphi} \partial^\mu \varphi) 
+ 12\e^{4\sigma-2\varphi}\partial_\mu\varphi \partial^\mu \varphi \nn
&& -2 \e^{2\sigma-2\varphi} \left[
-6\Box\sigma - 6 (\partial_\mu\sigma )
(\partial^\mu\sigma)\right] \ .
\eea
Since we are interested in cosmological problems, we only consider the
uniform solution, where all the fields depend only on (conformal) time 
$t$ and we obtain
\bea
\label{sigmaeq2}
0&=& 4b'{d^4 \sigma \over dt^4} \nn 
&& - 2[3b'' + 2 (b + b')]\left[
{d^2 \over dt^2} \left[
{d^2 \sigma \over dt^2} 
+ \left({d \sigma \over dt} \right)^2\right]
-2{d \over dt}\left\{{d \sigma \over dt}
\left[{d^2 \sigma \over dt^2} 
+ \left({d \sigma \over dt} \right)^2\right]\right\}\right] \nn
&& +{a_1 \over 8}
\left({d \varphi \over dt} \right)^4  
+ {a_2 \over 4}
{d^2 \over dt^2} \left(\left({d \varphi \over dt} \right)^2 \right)  
- {a_2 \over 2}{d \over dt}
\left( {d \sigma \over dt}\left({d \varphi \over dt} \right)^2 \right) \nn
&& - 24 \e^{4\sigma-2\varphi}\left({d \varphi \over dt} \right)^2 
+ 2\e^{2\sigma-2\varphi} 
\left[-6{d^2 \sigma \over dt^2} 
-6 \left({d \sigma \over dt} \right)^2\right] \nn
&& -6 {d^2 \e^{2\sigma-2\varphi} \over dt^2}
+ 12 {d \over dt} \left(\e^{2\sigma-2\varphi}
{d \sigma \over dt} \right) \ . \\
\label{vpeq2}
0&=& {a_1 \over 2} {d \over dt} \left\{ \left(
{d \varphi \over dt}\right)^3 \sigma \right\} 
- {a_2 \over 2}{d  \over dt}
\left({d \varphi \over dt}{d^2 \sigma \over dt^2} \right) 
- {a_2 \over 2}{d  \over dt}
\left({d \varphi \over dt}\left({d \sigma \over dt}\right)^2 \right) \nn
&& + 12 {d \over dt} \left( \e^{4\sigma -2\varphi} 
{d \varphi \over dt} \right) 
+ 12\e^{4\sigma-2\varphi}\left({d \varphi \over dt}\right)^2 \nn
&& -2 \e^{2\sigma-2\varphi} 
\left[-6{d^2 \sigma \over dt^2} 
-6 \left({d \sigma \over dt} \right)^2\right] \ .
\eea
We can often drop the terms linear to $\sigma$ in (\ref{v13}) \cite{BOS}.
(Actually, if one considers the correspondent theory as IR sector of 
quantum gravity \cite{A} 
or IR sector of induced supergravity
\cite{BU} the omitting of linear term on sigma corresponds 
to trivial redefinition of the correspondent source.)
In this case, (\ref{sigmaeq2}) and (\ref{vpeq2}) can be reduced to 
\bea
\label{sigmaeq3}
0&=& 4b'{d^4 \sigma \over dt^4} \nn 
&& - 2[3b'' + 2 (b + b')]\left[
{d^2 \over dt^2} \left[
{d^2 \sigma \over dt^2} 
+ \left({d \sigma \over dt} \right)^2\right]
-2{d \over dt}\left\{{d \sigma \over dt}
\left[{d^2 \sigma \over dt^2} 
+ \left({d \sigma \over dt} \right)^2\right]\right\}\right] \nn
&& - {a_2 \over 2}{d \over dt}
\left( {d \sigma \over dt}\left({d \varphi \over dt} \right)^2 \right) \nn
&& - 24 \e^{4\sigma-2\varphi}\left({d \varphi \over dt} \right)^2 
+ 2\e^{2\sigma-2\varphi} 
\left[-6{d^2 \sigma \over dt^2} 
-6 \left({d \sigma \over dt} \right)^2\right] \nn
&& -6 {d^2 \e^{2\sigma-2\varphi} \over dt^2}
+ 12 {d \over dt} \left(\e^{2\sigma-2\varphi}
{d \sigma \over dt} \right) \ . \\
\label{vpeq3}
0&=& - {a_2 \over 2}{d  \over dt}
\left({d \over dt}\left({d \sigma \over dt}\right)^2 \right) \nn
&& + 12 {d \varphi \over dt} \left( \e^{4\sigma -2\varphi} 
{d \varphi \over dt} \right) 
+ 12\e^{4\sigma-2\varphi}\left({d \varphi \over dt}\right)^2 \nn
&& -2 \e^{2\sigma-2\varphi} 
\left[-6{d^2 \sigma \over dt^2} 
-6 \left({d \sigma \over dt} \right)^2\right] \ .
\eea
In the large $N$ limit, where the contribution from the classical part can 
be neglected in Eqs.(\ref{sigmaeq}) and (\ref{vpeq}), some analytical 
solutions can be found.  When the terms linear to $\sigma$ in (\ref{v13}) 
are dropped, such analytical solutions were found in \cite{OOG}.  On the other 
hand, when we include the $\sigma$ linear terms, we can construct only a few 
closed form solutions. Some of them are given by
\bea
\label{sol4d1}
(1)& \ \varphi=\varphi_0\ (\mbox{constant}),\ \ 
\sigma=0 & \nn
(2)& \ \varphi=\varphi_0\ (\mbox{constant}),\ \ 
\sigma=\alpha\ln\left({t \over t_0}\right),\ \  &
\alpha^2\equiv 1 - {2b' \over 3b''+ 2(b+b')} >1  \nn
(3)& \varphi=\beta \ln\left({t \over t_0}\right),\ \  
\sigma=0,\ \ &
\beta^2\equiv 12{a_2 \over a_1} >0
\eea
In the case that the terms linear to $\sigma$ in (\ref{v13}) are dropped 
\cite{OOG} in the large $N$ limit, (\ref{sigmaeq3}) and (\ref{vpeq3}) can 
be integrated to be
\bea
\label{O8}
&& r{d^3\sigma \over dt^3}-2q\left({d\sigma \over dt}\right)^3
-{a_2 \over 4}{d \over dt}\left(\left({d\varphi \over dt}
\right)^2\right)=c \nn
&& {d \varphi \over dt}\left({d\sigma \over dt}\right)^2=e
\eea
where $c$ and $e$ are integration constants and $r=-(3b''+2b)$, $q=-[3b''+
2b'+2b]$.  We can find some particular solutions of Eqs. (\ref{O8}).
\begin{enumerate}
\item For $r=0$, we get 
\bea
\label{O10}
\sigma&=&\sigma_0 t+ \mbox{const} \nn
f&=&\mbox{const} \exp\left(\frac{et}{\sigma_0^2}\right)
\eea
where
\be
\label{O10b}
\sigma_0=\left(-\frac{a_2e^2}{4q}
\pm\sqrt{\frac{ae^2}{4q}-\frac{c}{2q}}\right)^{1 \over 3}
\ee
That is a singular solution, it has an initial singularity at 
physical time equal to zero.\\
\item For $r\ne 0$, we find the general solution:
\be
\label{O11}
\sigma=\int \frac{y dy}{\tilde{u}(y)}\ ,\ \ 
t=\int \frac{dy}{\tilde{u}(y)}\ ,\ \ 
\varphi =-{e \over 2}\int \frac{dy}{y^2\tilde{u}(y)}
\ee
where
\be
\label{O11b}
\tilde{u}^2(y)=\frac{1}{r}\left(qy^4-\frac{a_2e^2}{y^2}+2cy+e\right)
\ee
One particular solution in (\ref{O11b}) is given by
\be
\label{O11c}
\varphi =-{e \over 2}t\ ,\ \ \sigma=t\ , \ \ 
c=-2q-a_2e^2
\ee
That is the same solution as in (\ref{O10}). 
Another particular solution is given by $q=0$, $c=0$, $e=1$, $r=2b'$, 
and
\bea
\label{O11d}
\sigma&=&a_2\sqrt{|2b'|}\left[\frac{1}{2}
\arcsin\left(\frac{(|2b'|)^{-1/2}t +a_2}
{a_2}\right)^{1/2}\right.\nn
&& \left.+\frac{1}{4}\sin\left\{
2\arcsin\left(\frac{(|2b'|)^{-1/2}t +a_2}{a_2}\right)^{1/2}
\right\}\right],
\eea
and similar solution for $\varphi$.
That is again singular solution.
In general, dilatonic coupling acts against such non-singular 
solutions. We did not find any non-singular solutions 
from the general solution (\ref{O11}).
\end{enumerate}

We now consider solving (\ref{sigmaeq2}) and (\ref{vpeq2}) or 
(\ref{sigmaeq3}) and (\ref{vpeq3}) numericaly.  Some examples 
of the solution by numerical calculations are given in Figs.11 -- 14. 
The results in case that we include the terms in (\ref{v13}) linear 
to $\sigma$ (\ref{sigmaeq2}) and (\ref{vpeq2}) are almost identical 
with the case that we drop the terms linear to $\sigma$ (\ref{sigmaeq3}) 
and (\ref{vpeq3}), as expected. For this reason, we only give the 
results in Figs.11 -- 14 for the case that we drop the terms linear 
to $\sigma$.  In Figs.11 -- 14, we calculate $\phi$ (solid line), 
$\sigma$ (broken line) and 4d scalar curvature (dotted line), which 
is given by
\be
\label{4dscv}
R=-6{\partial^2 \sigma \over \partial t^2}
-6 \left( {\partial \sigma \over \partial t} \right)^2\ ,
\ee
for several $N$ and for the several initial conditions.  An interesting 
thing is that the 4d scalar curvature oscillates as a trigonometric 
function. In Fig.11, we choose $N=1$ and the initial condition 
$\phi(0)=\sigma(0)=0$, $\dot\phi(0)=\dot\sigma(0) =\ddot\sigma(0)=
\sigma^{(3)}(0)=1$. In Fig.11, both of the dilaton field $\phi$ and the 
conformal factor $\sigma$ slowly increase monotonically but the scalar 
curvature vibrates rather quickly like trigonometric functions but the 
amplitude slowly increases.  In Fig.12, where we chose $N=100$ and the 
initial condition $\phi(0)=\sigma(0)=\sigma^{(3)}(0)=0$, $\dot\phi(0)=
\dot\sigma(0)=\ddot\sigma(0)=1$, the conformal factor $\rho$ behaves in
a rather complicated way, i.e., $\rho$ increases at first but decreases 
after that with slow and small oscillations. On the other hand, $\phi$ 
increases monotonically and reaches a singularity in finite conformal 
time.  Since the conformal factor $\rho$ does not become positive infinity, 
the finite conformal time corresponds to finite cosmological time.
The 4d scalar curvature vibrates  and the amplitude increases quickly 
near the singularity.  The 4d scalar curvature oscillates slowly in Fig.13, 
where $N=100$ and the initial condition is $\phi(0)=\sigma(0)=\dot\sigma(0)
=\ddot\sigma(0)=\sigma^{(3)}(0)=0$, $\dot\phi(0)=1$. The conformal factor 
$\rho$ increases at first but decreases after that with very slow 
oscillations and $\phi$ increases monotonically and reaches a singularity 
in finite conformal and cosmological times.  In Fig.14, where $N=100$ with 
the initial condition $\phi(0)=\dot\phi(0)=\sigma(0)=\ddot\sigma(0)=
\sigma^{(3)}(0)=0$, $\dot\sigma(0)=1$, the scalar curvature oscillates
very quickly and the period becomes short with the passing of conformal 
time. In this solution, the dilaton field $\phi$ also oscillates but 
the amplitude becomes small and the conformal factor $\rho$ slowly and 
monotonically increase. 

Hence, unlike the case of pure Einstein gravity with conformal matter 
back-reaction on gravitational background (see ref.\cite{S}) it is 
more difficult to realise the inflationary Universe in the theory 
with dilaton under consideration.

\section{Discussion}.

~~~In the present work we studied quantum cosmology for some 2d 
and 4d dilatonic SGs with dilaton coupled matter. Working in the large $N$ 
approximation we demonstrated that quantum matter back reaction does not really  
help to solve the singularity problem unlike the case with no external 
dilaton. Of course, this does not mean that we claim that the singularity 
problem cannot be solved as result of quantum fluctuations. Rather our 
results indicate that the solution of this problem perhaps could be possible 
only in complete quantum SG models, i.e. taking into account quantum effects 
of all fields (graviton,dilaton,gravitino,...). It would be really interesting 
to study such a problem but it is not easy to do since we do not know the 
correspondent expression for effective action (at best, it is known only 
as an expansion in the curvature,see \cite{BOS} for a 
review). Note, nevertheless,
that we found few non-singular Universes which are non-singular at early times 
but become singular at late times (or vice-versa).

Another interesting aspect of our work is the following one. If we exchange 
the role of time and radius for many of our numerical solutions we find a 
static object which may be interpreted as a 2d black hole (BH). 
A few years ago 
there was a lot of activity in the study of quantum dilatonic 2d BHs (for 
very incomplete list of references see \cite{RST} and refs. therein). Hence, 
using our approach one can study 2d dilatonic BHs for the models of dilaton 
SG with quantum matter.  It is also interesting to note that 
recently discussed 
effect of BH anti-evaporation \cite{BH1} also takes place in above 
BHs realized 
in SGs as it could be confirmed by easy calculation. (It is natural as only 
the coefficient of the Polyakov term in the anomaly induced effective 
action is changed 
due to the fermion's contribution).

As one more possible generalization of this work one can consider not only 
conformally flat but more complicated FRW cosmologies for 4d theory. However,
we do not expect that the results will be qualitatively very different at least 
in the models of the same sort. The reason is that main effects come through 
time dependence which will be similar.

\ 

\noindent
{\bf Acknoweledgments}
The work by SDO has been partially supported by Universidad del Valle and 
COLCIENCIAS(COLOMBIA).The research of SJG is supported by the U.S. National
Science Foundation under grant PHY-96-43219.


\newpage
\noindent
{\bf Figure Captions}

\ 

\noindent
Fig.1 $\phi$ (solid line), $\rho$ (broken line) 
and $R/100$ (dotted line) for $N=1$, $A^2=1$ and $\lambda^2=1$.

\noindent
Fig.2 $\phi/10$ (solid line), $\rho$ (broken line) 
and $R/100$ (dotted line) for $N=10$, $A^2=0$ 
and $\lambda^2=1$.

\noindent
Fig.3 $\phi$ (solid line), $\rho$ (broken line) 
and $R/30$ (dotted line) for $N=10$, $A^2=1$ 
and $\lambda^2=0$.

\noindent
Fig.4 $\phi$ (solid line), $\rho$ (broken line) 
and $R/500$ (dotted line) for $N=100$, $A^2=1$ and $\lambda^2=1$.

\noindent
Fig. 5 $\phi$ (solid line), $\rho$ (broken line) 
and $R_4/100$ (dotted line) for $N=1$, 
$A^2=0$ and $\lambda^2=1$.

\noindent
Fig. 6 $\phi$ (solid line), $\rho$ (broken line) 
and $R_4/100$ (dotted line) for 
$N=10$, $A^2=1$ and $\lambda^2=0$.

\noindent
Fig.7 $\phi$ (solid line), $\rho\times 10$ (broken line) 
and $\theta$ (dotted line) for $N=1$, $A^2=0$ and $\Lambda^2=1$.

\noindent
Fig.8 $\phi$ (solid line), $\rho\times 4$ (broken line) 
and $\theta$ (dotted line) for $N=1$, $A^2=1$ and $\Lambda^2=0$.

\noindent
Fig.9 $\phi\times 5$ (solid line), $\rho$ (broken line) 
and $\theta$ (dotted line) for $N=10$, $A^2=1$ and $\Lambda^2=1$.

\noindent
Fig.10 $\phi\times 100$ (solid line), $\rho\times 10$ (broken line) 
and $\theta$ (dotted line) for $N=100$, $A^2=1$ and $\Lambda^2=0$.

\noindent
Fig.11 $\phi$ (solid line), $\sigma$ (broken line) and 
4d curvature divided by $100$ (dotted line) for $N=1$ with the initial 
condition $\phi(0)=\sigma(0)=0$, $\dot\phi(0)=\dot\sigma(0)
=\ddot\sigma(0)=\sigma^{(3)}(0)=1$.

\noindent
Fig.12 $\phi$ (solid line), $\sigma\times 5$ (broken line) and 
4d curvature divided by $10$ (dotted line) for $N=100$ with the initial 
condition $\phi(0)=\sigma(0)=\sigma^{(3)}(0)=0$, 
$\dot\phi(0)=\dot\sigma(0)=\ddot\sigma(0)=1$.

\noindent
Fig.13 $\phi$ (solid line), $\sigma\times 10$ (broken line) and 
4d curvature divided by $10$ (dotted line) for $N=100$ with the initial 
condition $\phi(0)=\sigma(0)=\dot\sigma(0)=\ddot\sigma(0)=
\sigma^{(3)}(0)=0$, $\dot\phi(0)=1$.

\noindent
Fig.14 $\phi\times 500$ (solid line), $\sigma$ (broken line) and 
4d curvature divided by $5$ (dotted line) for $N=100$ with the initial 
condition $\phi(0)=\dot\phi(0)=\sigma(0)=\ddot\sigma(0)=
\sigma^{(3)}(0)=0$, $\dot\sigma(0)=1$.

\

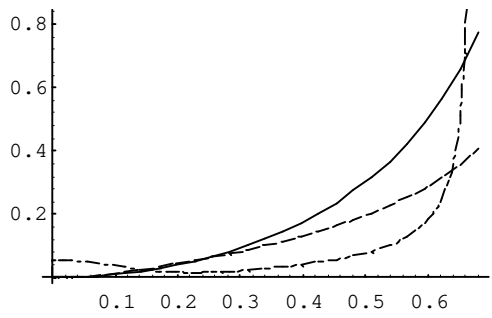
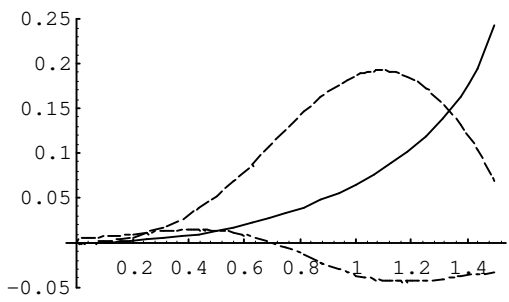
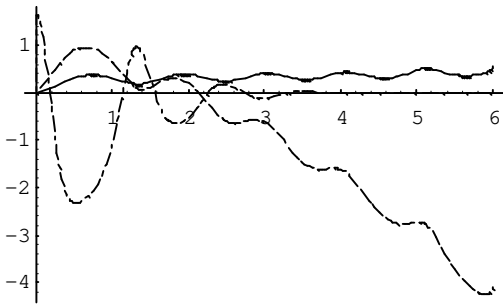
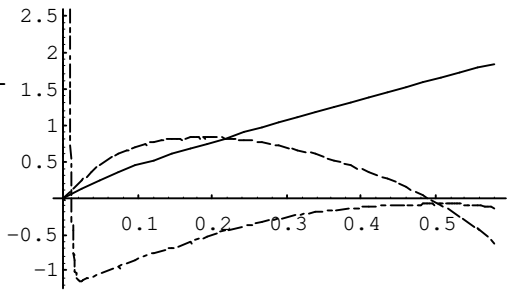
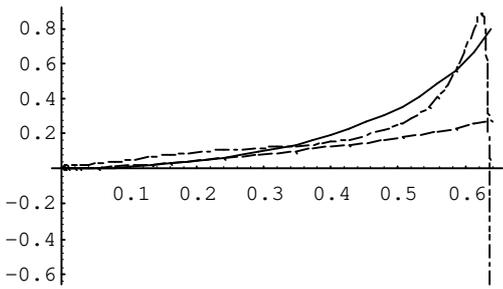
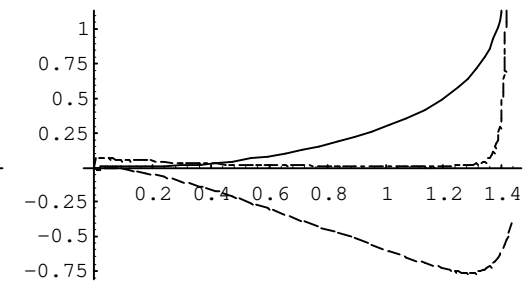
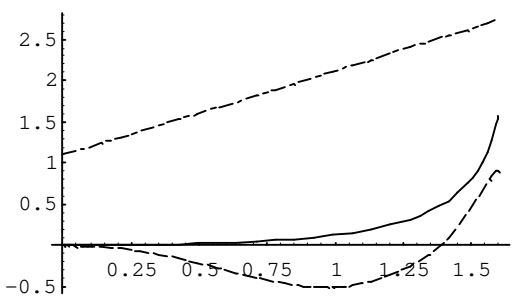
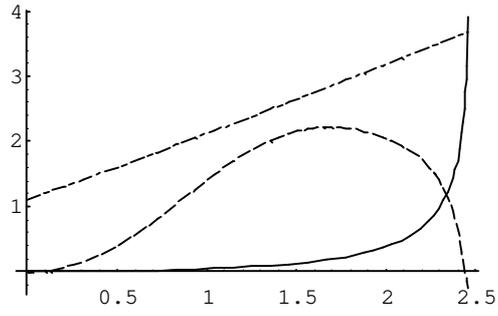
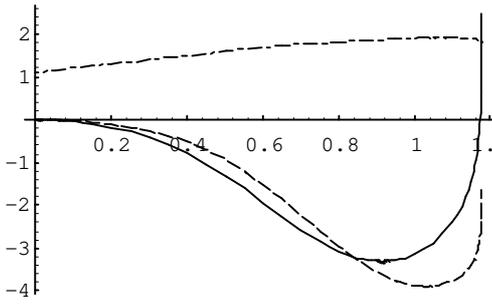
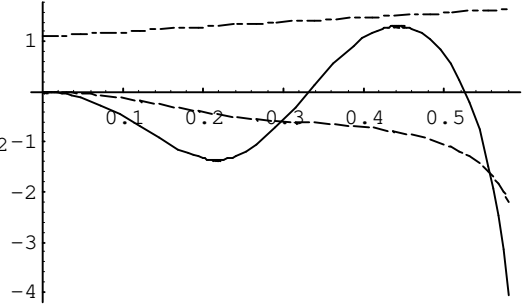
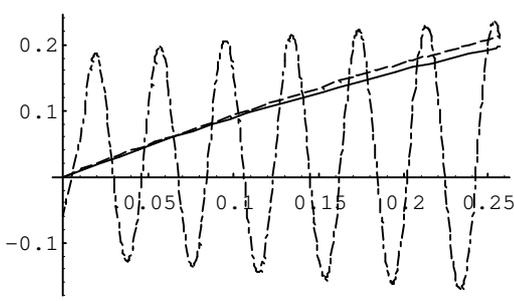
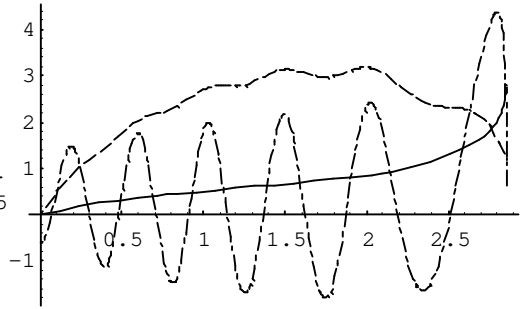
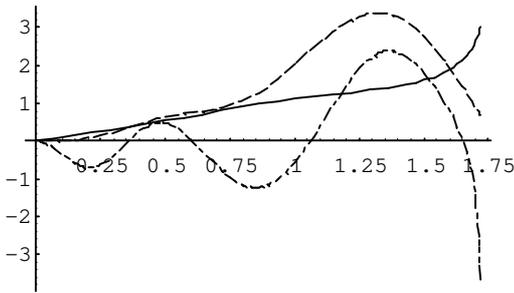
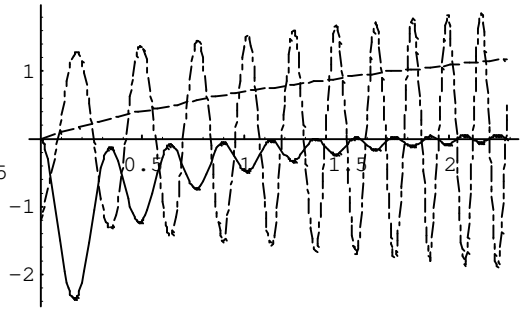
\end{document}